\documentclass[%
 superscriptaddress,
 reprint,
 showpacs,preprintnumbers,
 nofootinbib,
 amsmath,amssymb,
 aps,
 prd,
 floatfix,
]{revtex4-1}
\usepackage{graphicx}
\usepackage{placeins}
\usepackage{float}
\usepackage{xcolor}
\usepackage[normalem]{ulem}
\usepackage{hyperref}
\hypersetup{
    colorlinks=true,
    linkcolor=blue,
    filecolor=blue,      
    urlcolor=blue,
    pdftitle={CRPAinGENIE},
    }

\newcommand{\be}{\begin{equation}}
\newcommand{\ee}{\end{equation}}

\usepackage{lineno}

\begin{document}

\title{Implementation of the CRPA model in the GENIE generator and an analysis of nuclear effects in low-energy transfer neutrino interactions}

\author{S.~Dolan}
\affiliation{CERN, European Organization for Nuclear Research, Geneva, Switzerland}
\email[Contact e-mail: ]{Stephen.Joseph.Dolan@cern.ch}

\author{A.~Nikolakopoulos}
\affiliation{Fermi National Accelerator Laboratory, Batavia, IL 60510, USA}
\email[Contact e-mail: ]{anikolak@fnal.gov}

\author{O.~Page}
\affiliation{School of Physics, University of Bristol, Bristol BS8 1TL, United Kingdom}

\author{S.~Gardiner}
\affiliation{Fermi National Accelerator Laboratory, Batavia, IL 60510, USA}

\author{N.~Jachowicz}
\affiliation{Department of Physics and Astronomy, Ghent University, Proeftuinstraat 86, B-9000 Gent, Belgium}

\author{V.~Pandey}
\affiliation{Fermi National Accelerator Laboratory, Batavia, IL 60510, USA}
\affiliation {Department of Physics, University of Florida, Gainesville, FL 32611, USA}

\date{\today}
\begin{abstract}

We present the implementation and validation of the Hartree-Fock continuum random phase approximation (HF-CRPA) model in the GENIE neutrino-nucleus interaction event generator and a comparison of the subsequent predictions to experimental measurements of lepton kinematics from interactions with no mesons in the final state. These predictions are also compared to those of other models available in GENIE. It is shown that, with respect to these models, HF-CRPA predicts a significantly different evolution of the cross section when moving between different interaction targets, when considering incoming anti-neutrinos compared to neutrinos and when changing neutrino energies. These differences are most apparent for interactions with low energy and momentum transfer. It is also clear that the impact of nucleon correlations within the HF-CRPA framework is very different than in GENIE's standard implementation of RPA corrections. Since many neutrino oscillation experiments rely on their input model to extrapolate between targets, flavours, and neutrino energies, the newly implemented HF-CRPA model provides a useful means to verify that such differences between models are appropriately covered in oscillation analysis systematic error budgets. 

\end{abstract}

\maketitle

\section{Introduction}\label{sec:intro}

Whilst accelerator-based neutrino oscillation experiments such as T2K~\cite{T2K:2011qtm}, NOvA~\cite{NOvA:2007rmc}, Hyper-Kamiokande~\cite{Abe:2018uyc} and DUNE~\cite{DUNE:2020ypp}, offer an unprecedented opportunity to explore fundamental physics, such as the neutrino mass ordering and Charge-Parity violation (CPV) in the lepton sector, their success relies on a detailed understanding of sub-to-few-GeV neutrino-nucleus interactions. Unfortunately, no existing interaction model is able to quantitatively describe available data, necessitating the application of large systematic uncertainties in model predictions~\cite{Alvarez-Ruso:2017oui}. The impact of these uncertainties on neutrino oscillation analyses are often mitigated through the use of a ``near'' detector, which is exposed to the unoscillated neutrino beam, to constrain the uncertainties on the oscillated event rate at a ``far'' detector. However, a neutrino interaction model is still usually required to extrapolate between the different neutrino energies, beam flavour compositions, kinematic acceptances and sometimes different target materials of the near and far detectors. It is equally crucial that models are able to reliably predict the asymmetry between neutrino and anti-neutrino cross sections, such that these differences are not mistaken for a source of CPV. It is therefore important that systematic uncertainties on neutrino interaction models are able to reliably cover the plausible variation of differences in neutrino interaction cross sections between neutrino energies, flavours, kinematics and targets. 

It has previously been shown that the Hartree-Fock (HF) mean-field model for charged-current quasi-elastic (CCQE) interactions with continuum random phase approximation (CRPA) corrections, developed by the Ghent group~\cite{Jachowicz:2002rr,Pandey:2014tza}, is successful in describing inclusive electron scattering data and predicts significantly different cross sections at low energy transfer compared to Fermi-gas models~\cite{Nikolakopoulos:2018sbo}. The CRPA corrections account for long-range correlations through the same effective Skyrme interaction used for the HF mean-field~\cite{RYCKEBUSCH1988237,RYCKEBUSCH1989694}. As detailed in Ref.~\cite{Jachowicz:2021ieb}, a CRPA treatment of correlations represents a complementary approach to the ab-initio calculations detailed in Ref.~\cite{Lovato:2020kba}. Whilst CRPA does not contain the contributions beyond the two-nucleon force which are present in the ab-initio calculation, the two-nucleon correlations are iterated to all orders (which is not the case for ab-initio) and so the set of diagrams accounted for is therefore different. It is additionally interesting to note that the treatment of final state interactions (FSI) in HF-CRPA, via a distortion of the outgoing nucleon wave function, leads to significantly different predictions for muon and electron neutrino cross sections at low energy transfers compared to widely used plane wave impulse approximation (PWIA) model which do not include FSI~\cite{Nikolakopoulos:2019qcr}.

In this paper we report the implementation of the HF-CRPA model in the GENIEv3 neutrino-nucleus interaction event generator~\cite{Andreopoulos:2009rq,Andreopoulos:2015wxa} and evaluate how else it differs from other available CCQE models. Particular focus is placed on how the predictions differ between different nuclear targets and between neutrino and anti-neutrino interactions within the low energy and momentum transfer region where nuclear effects are most relevant. The predictions from the HF-CRPA model are compared to those of SuSAv2~\cite{Gonzalez-Jimenez:2014eqa,Megias:2016lke} (implemented in GENIE in Ref.~\cite{Dolan:2019bxf}) as well as the Valencia group's Local Fermi Gas (LFG) model~\cite{Nieves:2011pp} with and without RPA corrections to account for nucleon correlations. Since the LFG-RPA and HF-CRPA approaches start from different nuclear ground state models, and in particular because the HF model already incorporates some description of nucleon correlations, it is to be expected the impact of RPA and CRPA corrections are quite different.

The models are also compared to data provided by two T2K measurements reporting the cross section of charged current meson-less (CC0$\pi$) final states from interactions on carbon and oxygen targets~\cite{T2K:2020jav} for incoming neutrinos and anti-neutrinos~\cite{T2K:2020sbd}. To make a complete comparison, a 2-particle 2-hole (2p2h) and pion absorption contribution must be added to the CCQE predictions. Although the 1p1h and 2p2h contributions would ideally be computed within a single consistent framework, we are limited by which models are available in GENIE and so for all cases the SuSAv2 MEC model~\cite{RuizSimo:2016rtu,RuizSimo:2016ikw} is used to compute the 2p2h contribution. For other channels the models of GENIE configuration \texttt{G18\_10b} are used, containing the Berger-Seghal single pion production~\cite{BSSPP2007} model in addition to more inelastic channels fed through the ``hN'' intranuclear cascade model~\cite{Dytman:2021ohr} to predict possible meson-less final states.  

The paper is structured to first show the implementation scheme of the HF-CRPA model in Sec.~\ref{sec:implementation}.
A broad comparison of inclusive model predictions is then made in Sec.~\ref{sec:inccomp}, including comparisons of the models to T2K measurements. Conclusions are drawn in Sec.~\ref{sec:conc}.

\section{Implementation scheme}\label{sec:implementation}

The implementation of neutrino interaction models in neutrino event generators requires a fast method of calculating the differential cross section given some set of outgoing particle kinematics (as is required for standard rejection sampling methods). In order for the implemented model to exactly reflect the microscopic theory on which it is based, the cross section would need to be available as a function of the set of kinematics to describe the entire final state (i.e. the fully \textit{exclusive} cross section must be calculable). For the case of a CCQE interaction (neglecting any additional nuclear emission from FSI), this would require the calculation of the five-dimensional differential cross section as a function of the outgoing lepton and nucleon kinematics. 
However, few microscopic models are able to reliably provide such exclusive cross sections without relying on the factorized form implied by the PWIA (and none that can are currently implemented in neutrino event generators). Many models, such as the SuSAv2 approach, are specifically designed to provide \emph{inclusive} cross sections, yielding results as a function of the outgoing lepton kinematics only.
Such models can be implemented in generators using a ``factorisation approach'' (as detailed in~\cite{Dolan:2019bxf}) in which the lepton kinematics are calculated directly from a microscopic model calculation, before the hadronic system is added on top using approximate methods. For CCQE interactions, this typically involves: sampling a nucleon momentum and removal energy from some input spectral function; transferring it the appropriate four-momentum derived from the incoming neutrino energy and outgoing lepton kinematics; and then putting the resultant nucleon through a semi-classical FSI cascade model. Such an approach generally relies on the assumption that the nuclear ground state ``seen'' by the interaction is independent of the interaction's kinematics (although to partially alleviate this, it is possible to make the sampled spectral function depend on the four-momentum transfer). The resultant model can be seen to provide a fully accurate reflection of the microscopic models predictions for lepton kinematics but only a broad estimation for outgoing hadron kinematics given the information available. 
The HF-CRPA model that is the subject of this work is not explicitly limited to the calculation of inclusive observables, instead the energy and angle of the outgoing nucleon are obtained through a multipole decomposition~\cite{RYCKEBUSCH1988237,RYCKEBUSCH1989694}. One is still faced with the fact that, due to the presence of FSI in the distorted wave treatment, and further due to the RPA, the exclusive cross section does not factorise as in the PWIA. Retaining the full complexity of the model would thus require sampling in a higher-dimensional phase space, making the process inefficient. In the present work we hence only include the cross section in terms of lepton kinematics, by summing and integrating over the outgoing nucleon's energy and angle. The effect of the approximations made in the factorised approach described below can, in future work, be compared to a more complete implementation of the kinematic degrees of freedom, using for example the approach described in~\cite{Niewczas:2020fev}. This falls out of the scope of the present work however, where only the description of inclusive cross sections are considered.

The ``factorisation approach'' implementation scheme used to add HF-CRPA to GENIE is very similar to that used for the SuSAv2 CCQE model~\cite{Dolan:2019bxf}. The scheme benefits from the fact that the differential cross section can be written as the product of kinematic factors with the contraction of a generic lepton tensor and a model-specific hadron tensor, where the latter encodes all of the nuclear dynamics of the interaction. In this way, the implementation of HF-CRPA is achieved by inserting new hadron tensor look-up tables into GENIE (as previously done in~\cite{Dolan:2019bxf,Valencia2p2hInGenie:2016}). Separate tensors are provided for HF with and without CRPA corrections and for carbon, oxygen and argon targets.
Additionally separate tensors are provided for the charged-current neutrino and anti-neutrino interactions, which is a necessity for describing the cross section on asymmetric nuclei such as Argon. Small isospin breaking effects are also present in the responses even in the case of the even-even nuclei. These are due to the Coulomb potential of the nucleus which leads to differences in the binding energy of protons and neutrons in the initial state, but is also included consistently in the final-state for interactions in which a proton is emitted~\cite{Jachowicz:2021ieb}. The responses are further separated into their vector-vector, axial-axial, and vector-axial contributions. This separation makes it possible to consistently modify the axial form factor based on a single table by simply rescaling the axial-axial and vector-axial contributions. To predict cross sections for targets which do not have tables a simple ``scaling of the second kind'' is assumed~\cite{Amaro:2004bs,Megias:2017PhD}, extrapolating from the closest available tensor and accounting for the shift in the Fermi momentum between targets alongside an offset in the nuclear removal energy. Finally, further tensors are provided in which FSI is ``turned off'' via a replacement of the distorted nucleon wave function with a plane wave approximation. Whilst this is clearly unrealistic, this provides a means to study the potential impact of FSI effects on the inclusive cross-section model predictions.


The HF-CRPA model is especially well-suited to capture the non-trivial nuclear effects that manifest themselves at small energy and momentum transfers, for example the presence of giant-resonances. A non-uniform binning scheme for the hadron tables is therefore used to capture such fine details of the model, where the cross section evolves rapidly as function of the nuclear excitation energy due to such resonances.

The HF-CRPA calculations further provide a consistent treatment of long-range correlations beyond the HF mean field by using the same nucleon-nucleon interaction used to generate the HF mean field as residual interaction in the RPA. This interaction has to be regularized at large (four-)momentum transfers, as the Skyrme force is of zero range.
For this reason, in Ref.~\cite{Pandey:2014tza}, a cut-off in the residual interaction used in the CRPA was proposed. Its effect is that the CRPA cross section tends towards the HF result at large four momentum transfer in a way supported by an analysis of inclusive electron scattering data over a large kinematic region.
In Refs~\cite{Nikolakopoulos:2020alk,Jachowicz:2021ieb}, comparisons to $(e,e')$ data off heavier targets show that the full CRPA result provides a better description of the cross section for low (four-)momentum transfer, implying that this cut-off might be too strong at small values of momentum transfer $q$. 
The default behaviour of the present implementation is thus to provide the full CRPA result at small $q$ without the dipole cut-off. Through the extension of the cross section at large $q$, as detailed in Sec.~\ref{subsec:susainterp}, the results tend smoothly to the SuSAv2 result without the introduction of the cut-off in the residual interaction.

The hadron kinematics are calculated using identical methods to the SuSAv2 implementation, using a local Fermi gas spectral function with a custom momentum-transfer dependent removal energy derived from relativistic mean field model predictions~\cite{Dolan:2019bxf,Megias:2017PhD,Megias:2016ee,Gonzalez-Jimenez:2014eqa}. Validations of the model implementation are detailed and shown in Appendix~\ref{app:vali}.

\subsection{SuSAv2 Interpolation}
\label{subsec:susainterp}

The nuclear response obtained in HF-CRPA naturally evolves from the low-energy region into a robust description of the quasielastic regime~\cite{Pandey:2014tza}. However, as the nuclear responses within the model are not fully relativistic, its reliability might decrease with increasingly large (four-)momentum transfers. Tables are provided up to 1~GeV energy transfer and 2~GeV momentum transfer which may be used as-is, but the standard behaviour of the implementation is to interpolate between the responses calculated in HF/HF-CRPA\footnote{Throughout this article, ``HF/HF-CRPA'' is used to refer to the HF \textit{and} HF-CRPA models.} and SuSAv2 at large momentum transfer,  $q$.  

\begin{figure}[hbp]
\begin{center}
\includegraphics[width=0.48\textwidth]{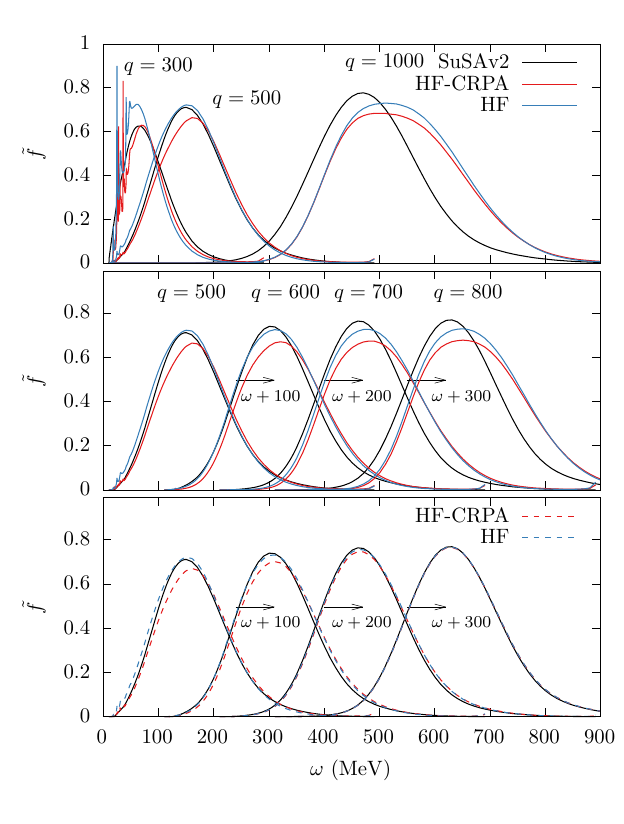}
\end{center}
\caption{The global scaling functions for charged-current quasielastic scattering of $\nu_e$ with $E_\nu=3~\mathrm{GeV}$ off oxygen within different models. The solid lines show the original HF and CRPA results. 
The dashed lines in the bottom panel shows the behaviour of the implementation where the HF and HF-CRPA cross sections are smoothly connected to the SuSAv2 result. 
In the middle and bottom panels the results are translated in $\omega$ as indicated for improved visibility.
}
\label{fig:SuSA_RPA_qdep}
\end{figure}

The interpolation is performed in a region of ``intermediate'' momentum transfers, where several approaches are found to give cross sections resembling the SuSAv2 results~\cite{Gonzalez-Jimenez:2019ejf}. 
This is illustrated in Fig.~\ref{fig:SuSA_RPA_qdep}, where the global scaling functions for $\nu_e$-induced charged-current ($CC$) interactions:
\begin{equation}
\tilde{f}^{CC}(E_\nu,q,\omega) = k_F\frac{\left[\frac{\,d\sigma^{CC}}{\,d\omega\,d\cos\theta_l}\right]}{\sigma_0^{CC}\left( v_L G^{CC}_L + v_T G^{CC}_T + v_{T^\prime} G^{CC}_{T^\prime} \right)},
\end{equation}
obtained with HF-CRPA are compared to those from the SuSAv2 parametrization at energies $E_\nu=3~\mathrm{GeV}$.
Here $\sigma_0^{CC} = \frac{G_F^2\cos\theta_c^2}{2\pi} k_l E_l$ and we use $k_F = 228~\mathrm{MeV}$. The single nucleon responses $G_i(\omega,q,k_F)$ and lepton factors $v_i$ are defined in Ref.~\cite{Gonzalez-Jimenez:2014eqa}. 
We show the scaling function rather than the differential cross section merely for benefits of presentation, as the $E_\nu$ and $q$-dependence mostly cancel in this way. 
Practically identical results are obtained for different $E_\nu > 1~\mathrm{GeV}$.

Note that the exact details of the SuSAv2 model used for the interpolation differ slightly from the usual GENIE implementation (reported in~\cite{Dolan:2019bxf}) such that there is a better correspondence with HF-CRPA at intermediate momentum transfers. In particular, the running removal energy and the RMF-RPWIA (RMF is the relativistic mean field model on which SuSAv2 is primarily based and RPWIA is a relativistic extension of PWIA) transition employed by SuSAv2~\cite{Gonzalez-Jimenez:2019ejf, Megias:2014qva}, both of which occur as as a function of momentum transfer, are slightly different in the GENIE implementation than in Ref.~\cite{Megias:2014qva}, the latter of which is used here.

One sees that for a value of $q\approx 500~\mathrm{MeV}$ the scaling functions obtained with HF/HF-CRPA are indeed similar to the SuSAv2 results.
For large $q$ the HF/HF-CRPA cross sections are shifted to larger $\omega$ compared to SuSAv2, in a similar way as the RMF results shown in Ref.~\cite{Gonzalez-Jimenez:2019ejf}.
The middle panel of Fig.~\ref{fig:SuSA_RPA_qdep} shows the cross section in the intermediate-$q$ region. The HF results again behave like the RMF, while the more peaked result of SuSAv2 is a result of a $q$-dependent combination of the aforementioned RMF and RPWIA scaling functions.

Motivated by these results, a fast transition is employed between the HF/HF-CRPA differential cross section at low-$q$ and SuSAv2 cross section at large $q$ in the region $ 500~\lesssim q \lesssim 700~\mathrm{MeV}$.
The interpolated cross section is parametrized as:
\begin{equation}
    \sigma = \cos\phi\left(q\right) \sigma^{HF-CRPA}+ \sin \phi\left(q\right) \sigma^{SuSAv2},
\end{equation}
where $\sigma$ stands for a double differential inclusive cross section and
\begin{equation}
\phi\left(q\right) = \frac{\pi}{2}\left[ 1 - \left(1+e^\frac{q-q_0}{\Delta_q}\right)^{-1} \right].
\end{equation}
The parameters $q_0 = 600~\mathrm{MeV}$ and $\Delta_q = 100~\mathrm{MeV}$, provide the fast transition required in a region where the SuSAv2 predictions are similar to the HF/HF-CRPA calculations. This interpolation is applied within the GENIE event generation where the differential cross sections are calculated at the energy of the event under consideration (there is no use flux averaged cross sections). The result of this procedure for the HF and HF-CRPA cross sections are shown in the lower panel of Fig.~\ref{fig:SuSA_RPA_qdep}.

\begin{figure}[htbp]
\begin{center}
\includegraphics[width=0.48\textwidth]{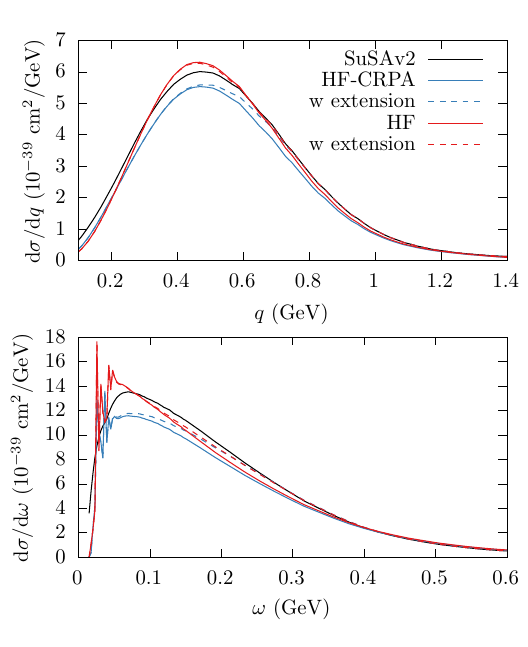}
\end{center}
\caption{A demonstration of the model extension employed between the HF-CRPA and SuSAv2 models. The single-differential T2K flux integrated cross section in terms of momentum and energy transfer $q$ and $\omega$ respectively are shown. The dashed lines show the results from HF-CRPA (red) and HF (blue) that include the interpolation to the SuSAv2 results, while the corresponding solid lines show the original models.
}
\label{fig:susaextension}
\end{figure}

Fig.~\ref{fig:susaextension} shows T2K flux-averaged cross sections with and without the extension with solid and dashed lines respectively.
Analysing the $q$-dependence in the top panel, in particular for the HF model, one sees that the cross section around and larger than the transition point of $600~\mathrm{MeV}$ is modified minimally. This indicates that, although the double-differential cross section has a different shape as seen in Fig.~\ref{fig:SuSA_RPA_qdep}, the total strength is similar (this is also seen by direct comparison of the total cross sections in Fig.~\ref{fig:sigmaenu}). It can be noted that the CRPA cross section with this extension grows in strength compared to the original model.

Whilst the choice of parameters is motivated by Fig.~\ref{fig:SuSA_RPA_qdep}, which shows calculations for neutrino interactions on an oxygen target, the model differences between SuSAv2 and HF/HF-CRPA were verified to remain very similar also for anti-neutrino interactions and for interactions on a carbon target. For interactions on argon the shape of the SuSAv2 prediction differs from HF/HF-CRPA a little more, but the chosen parameters still provide a reasonable interpolation. It is possible that future dedicated studies may be able to better tune the details of the interpolation but the impact of such tuning is expected to be very much a second order effect in dictating the model's predictions.

With the exception of the validation plots shown in Appendix~\ref{app:vali}, all predictions from the HF/HF-CRPA models shown employ this interpolation method. The SuSAv2 model shown is always the one from the original GENIE implementation which, as noted above, has small differences to the version used for the interpolation.

\section{Comparison of CRPA with other models in GENIE}\label{sec:inccomp}

\begin{figure}[b]
\begin{center}
\vspace{-5mm}
\includegraphics[width=0.52\textwidth]{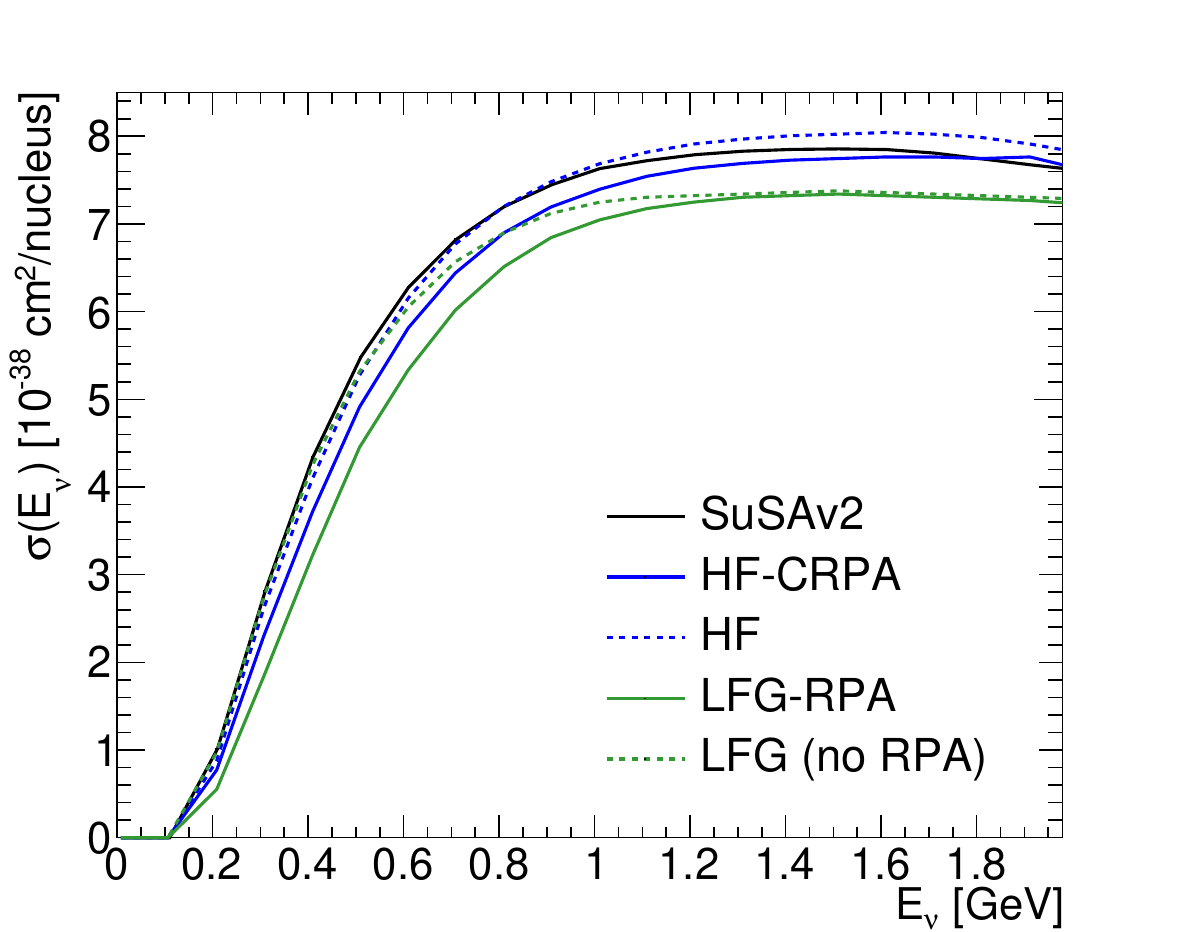} \vspace{-4mm}
\includegraphics[width=0.52\textwidth]{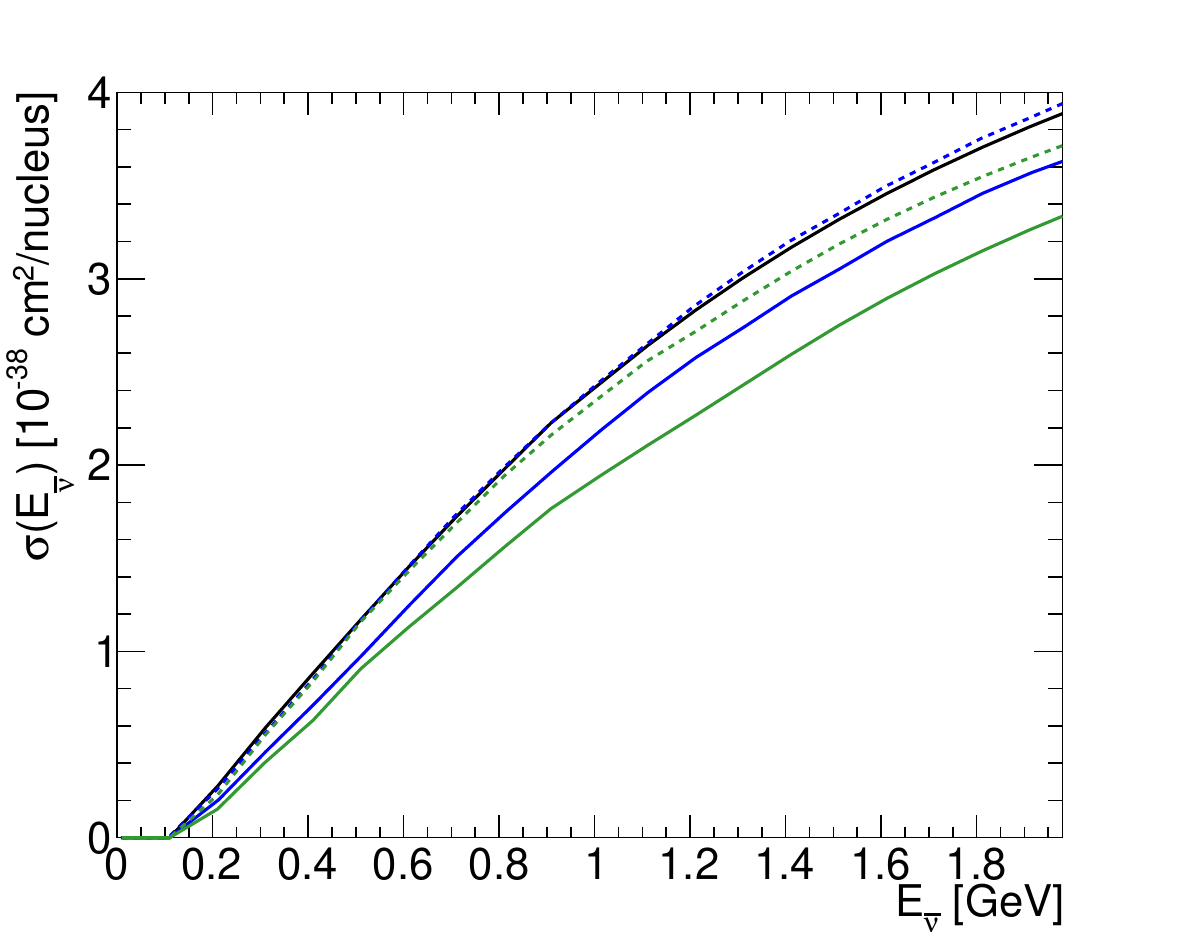} \vspace{-5mm}
\end{center}
\caption{The evolution of the total $\nu_\mu$ CCQE cross section predicted by various models as a function of neutrino energy is shown for neutrino and anti-neutrino cross sections on an oxygen target in the upper and lower plot respectively. 
}
\vspace{-3mm}
\label{fig:sigmaenu}
\end{figure}

\begin{figure*}[htbp]
\begin{center}
\includegraphics[width=0.32\textwidth]{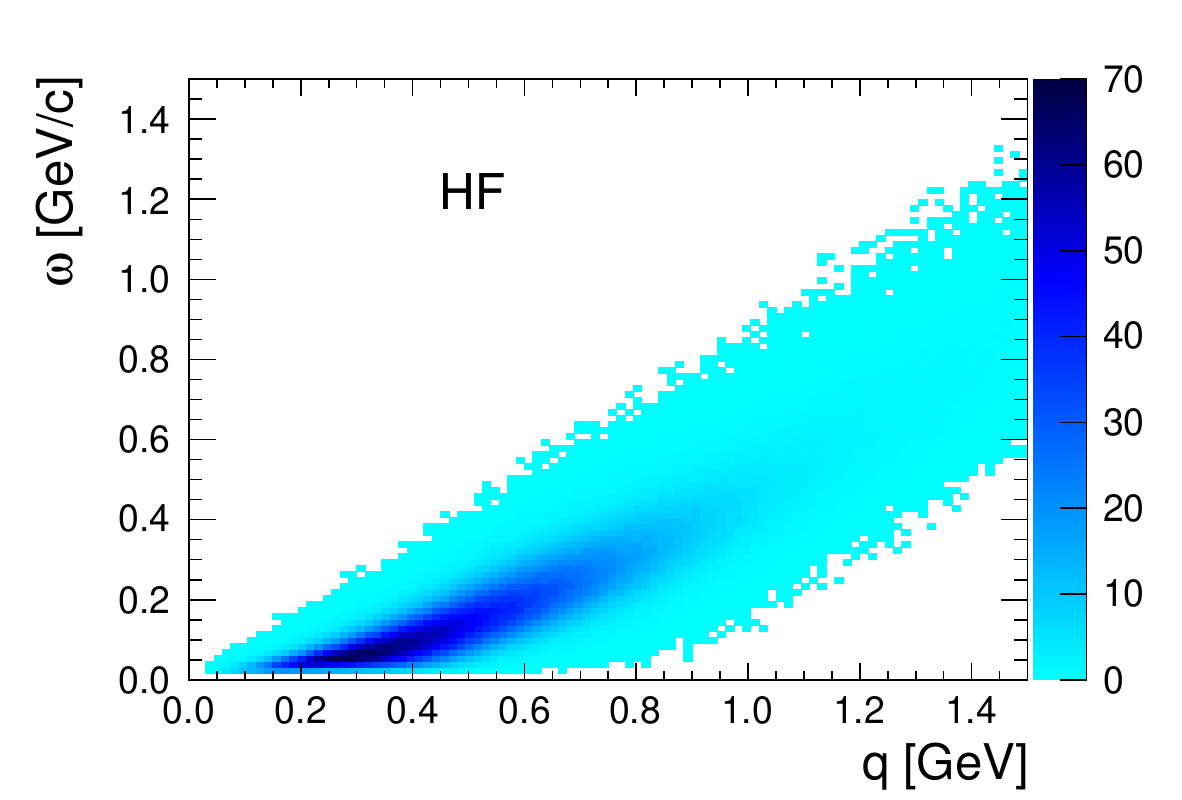}
\includegraphics[width=0.32\textwidth]{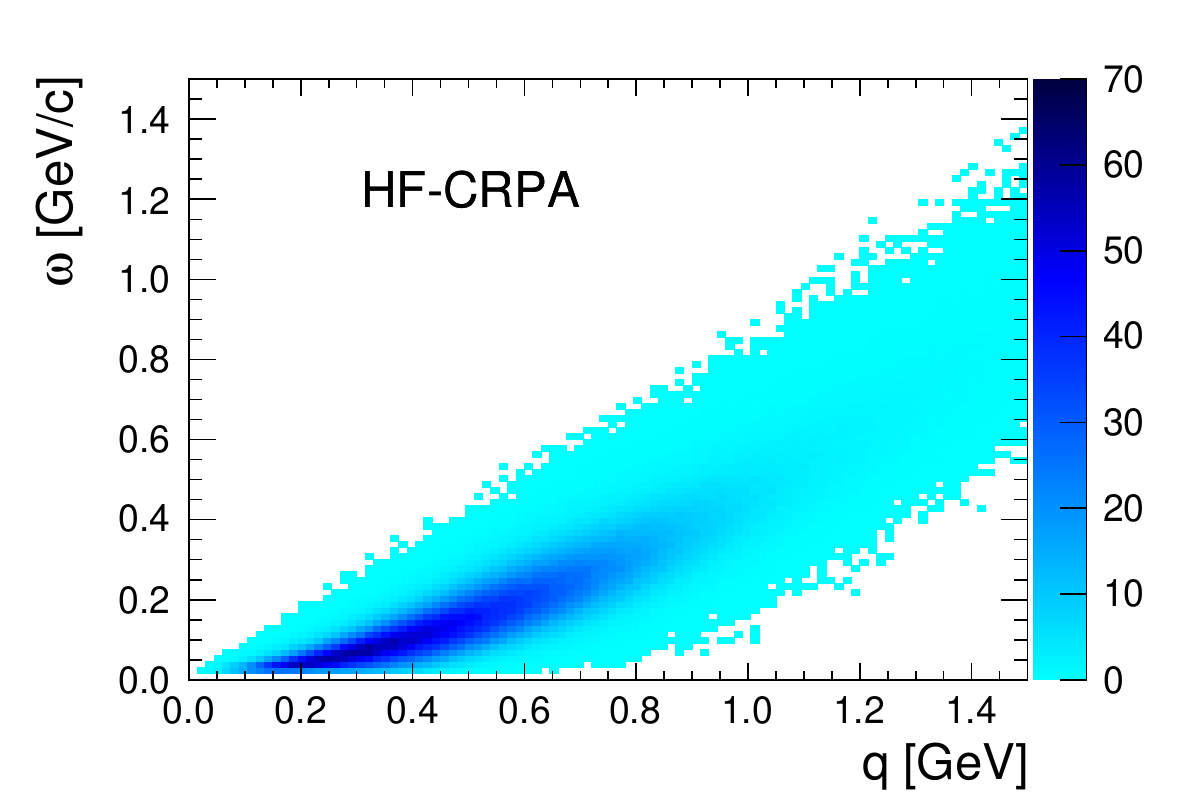}
\includegraphics[width=0.32\textwidth]{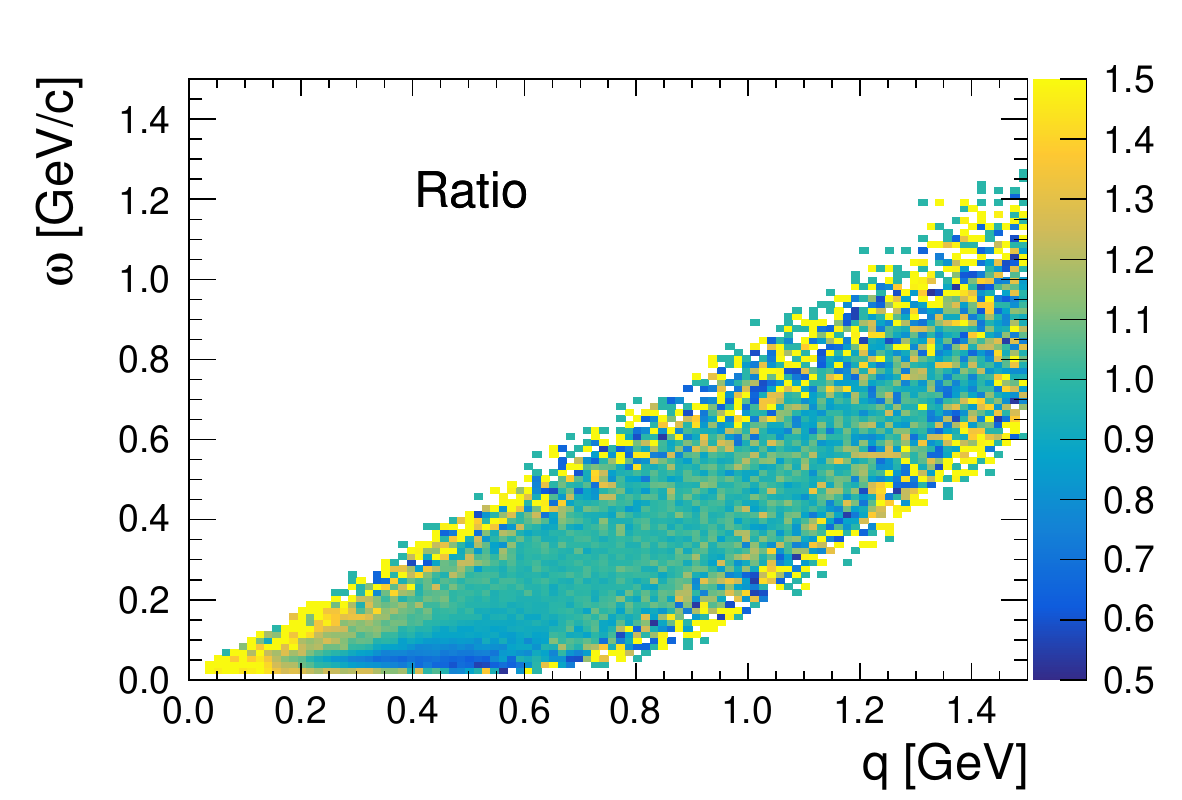}
\includegraphics[width=0.32\textwidth]{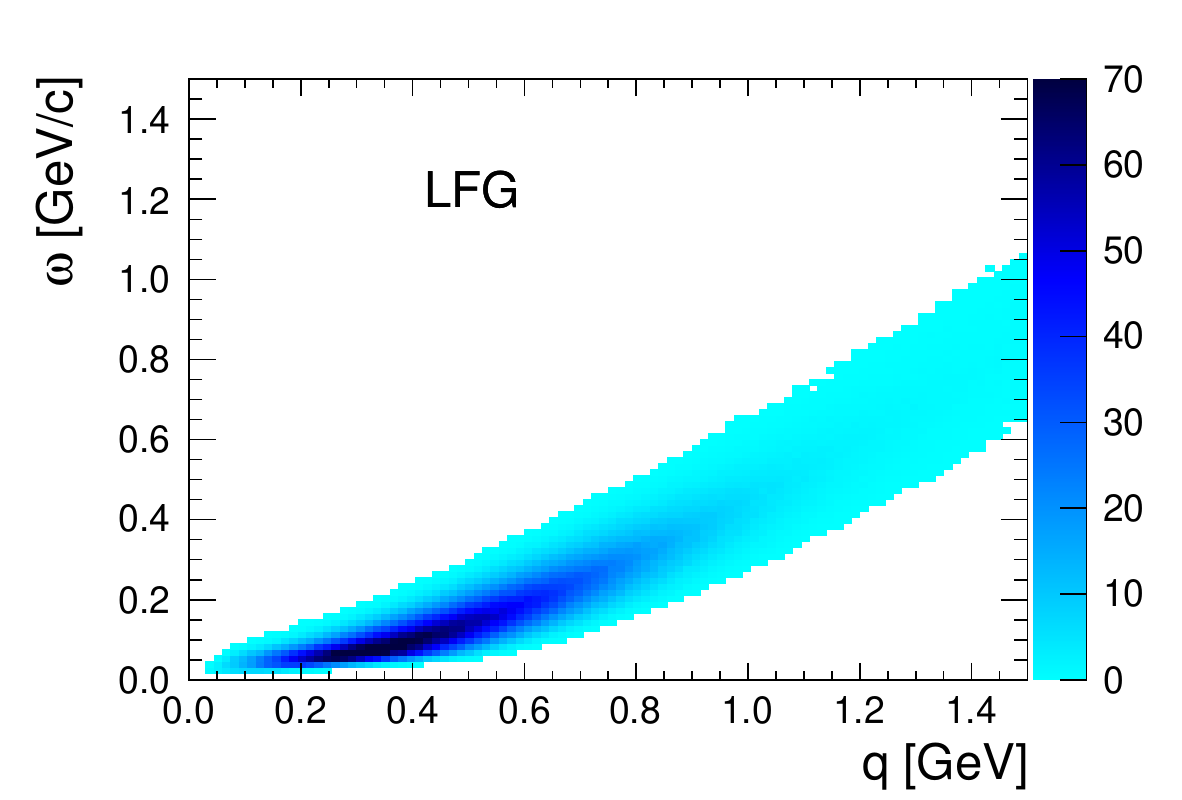}
\includegraphics[width=0.32\textwidth]{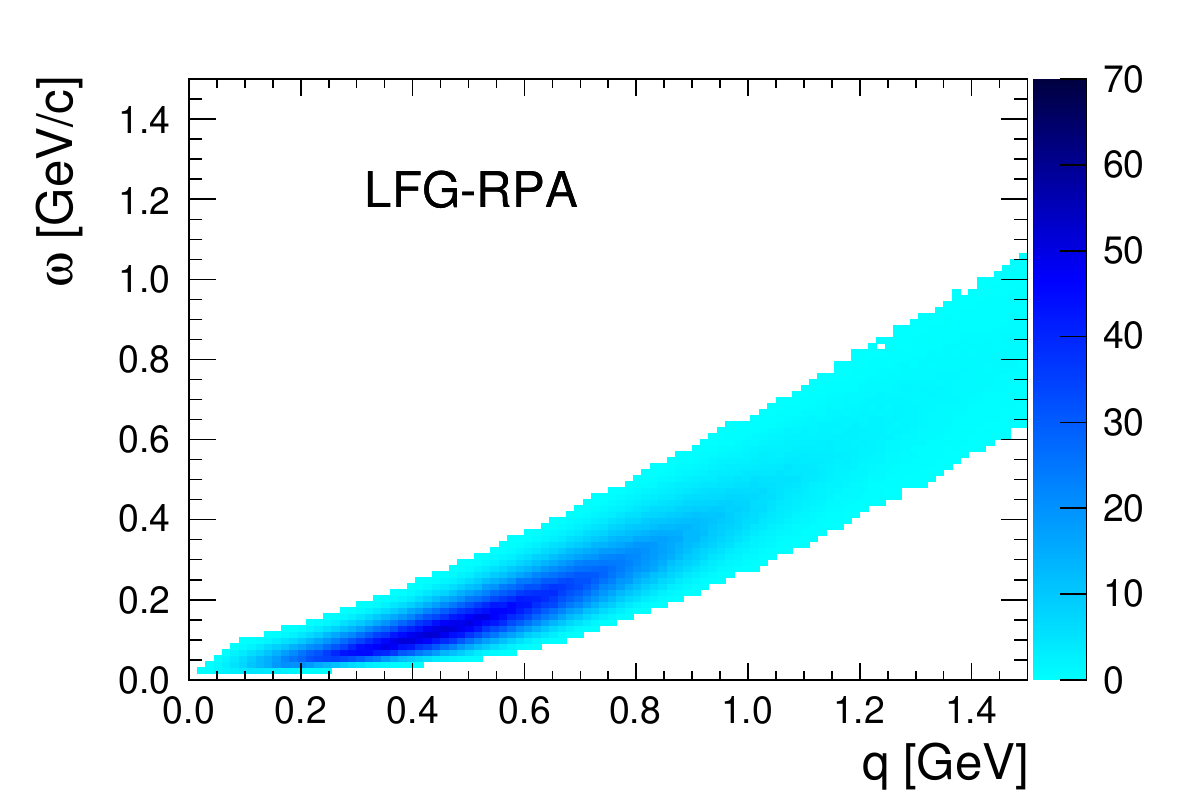}
\includegraphics[width=0.32\textwidth]{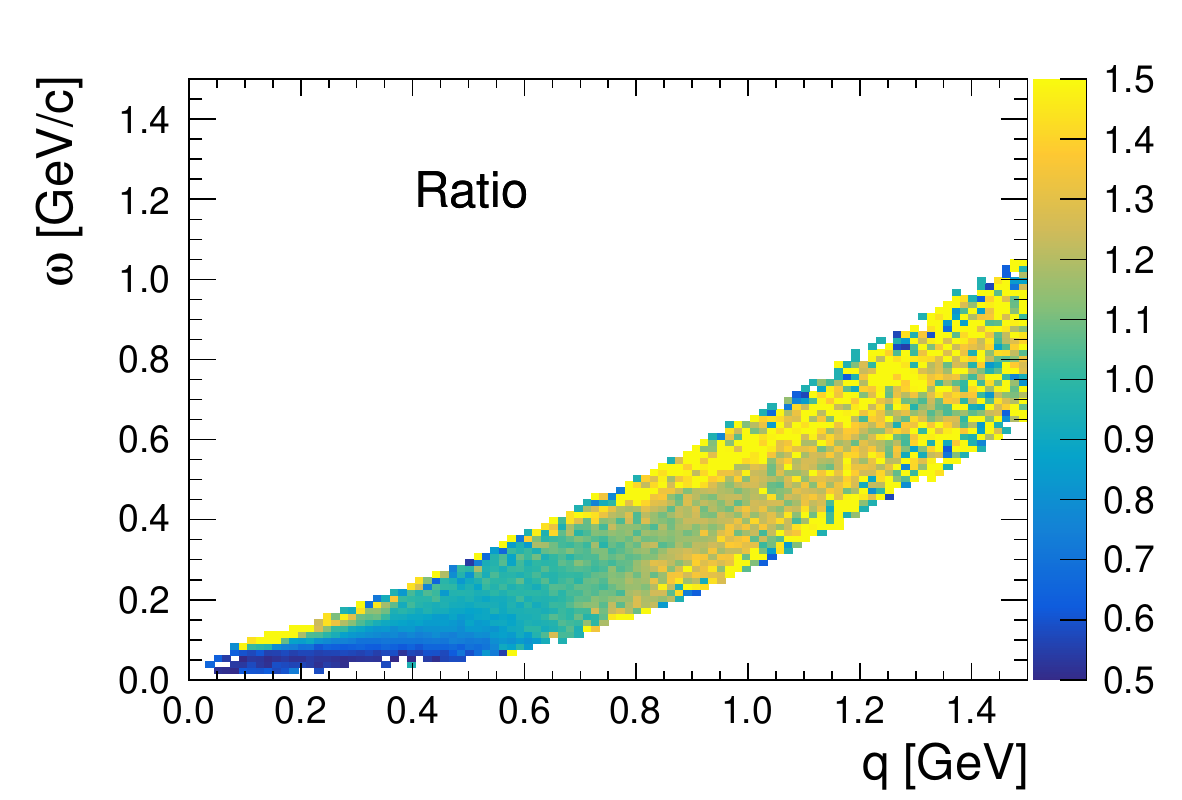}
\end{center}
\caption{The left and centre T2K muon neutrino flux (peaking at 0.6 GeV) averaged double differential cross section on carbon as a function of the energy ($\omega$) and momentum transfer ($q$) shown (on the z-axis) 10$^{-39}$ cm$^{2}$ GeV$^{-2}$/c per nucleon with and without (C)RPA corrections. All plots are produced with GENIE. The plot on the right shows the impact of the corrections as a ratio. The upper row shows predictions from HF/HF-CRPA model whilst the the lower row shows the same for the Valencia LFG/LFG-RPA model. White regions indicate regions in which no events were generated.}
\label{fig:q0q3}
\end{figure*}

The evolution of the neutrino and anti-neutrino cross sections as a function of neutrino energy for HF-CRPA as well as the other considered GENIE CCQE models are shown in Fig.~\ref{fig:sigmaenu}, while the double differential cross section as a function of energy ($\omega$) and momentum ($q$) transfer integrated over the T2K flux~\cite{Abe:2012av,t2kfluxurl} is shown in Fig.~\ref{fig:q0q3}. Note that the nucleon axial mass parameter within all models is set to between 0.99 GeV and 1.03 GeV. The cross section suppression from CRPA and RPA is clear in both figures, although it can be seen that the shape of the suppression and the differences in the neutrino and anti-neutrino cases are quite different. As discussed in Sec.~\ref{sec:intro}, it should be noted that the physics content of an RPA approach is determined by the residual interaction and the mean-field propagator, which are significantly different in the LFG-RPA and HF-CRPA approaches as discussed e.g. in~\cite{Martini:2016eec}. In particular, the ground state model in the HF-(CRPA) case (which is just HF) already includes a mean-field propagator, whilst LFG uses a free propagator, and so it is expected that the impact of CRPA on top of HF is quite different than RPA on top of LFG. 

In general the suppression from RPA in the Valencia group's LFG model is concentrated at low $q$ and causes a small enhancement of the cross section at larger $\omega$ and $q$. CRPA instead causes very little enhancement of the cross section in any region of kinematic phase space and its suppression is generally significantly weaker. It can also be noted that CRPA's suppression acts most strongly at slightly higher $\omega$ compared to the RPA case and is confined to a tighter region of $\omega$, $q$. Fig.~\ref{fig:q0q3} additionally demonstrates that the GENIE implementation of the Valencia LFG model is restricted to producing events within a limited kinematic phase-space close to the peak region, whilst HF-CRPA is not. 

The lower $\omega$, $q$ region in which RPA impacts GENIE's LFG model also manifests as a much stronger suppression than that caused by CRPA on top of HF at low neutrino energy (before the cross section saturates and so when the low $\omega$, $q$ is responsible for a larger portion of the cross section). In general the relative size of the suppression is larger for CRPA at higher neutrino energies and for RPA at lower neutrino energies. Since anti-neutrino interactions have a larger portion of their cross section at lower energy transfers compared to their neutrino counterparts~\cite{Pandey:2013cca} (due to the sign of the transverse interference term in the cross section), the impact of RPA continues to act as a significant suppression up to larger anti-neutrino energies.

\begin{figure*}[htbp]
\begin{center}
\includegraphics[width=0.48\textwidth]{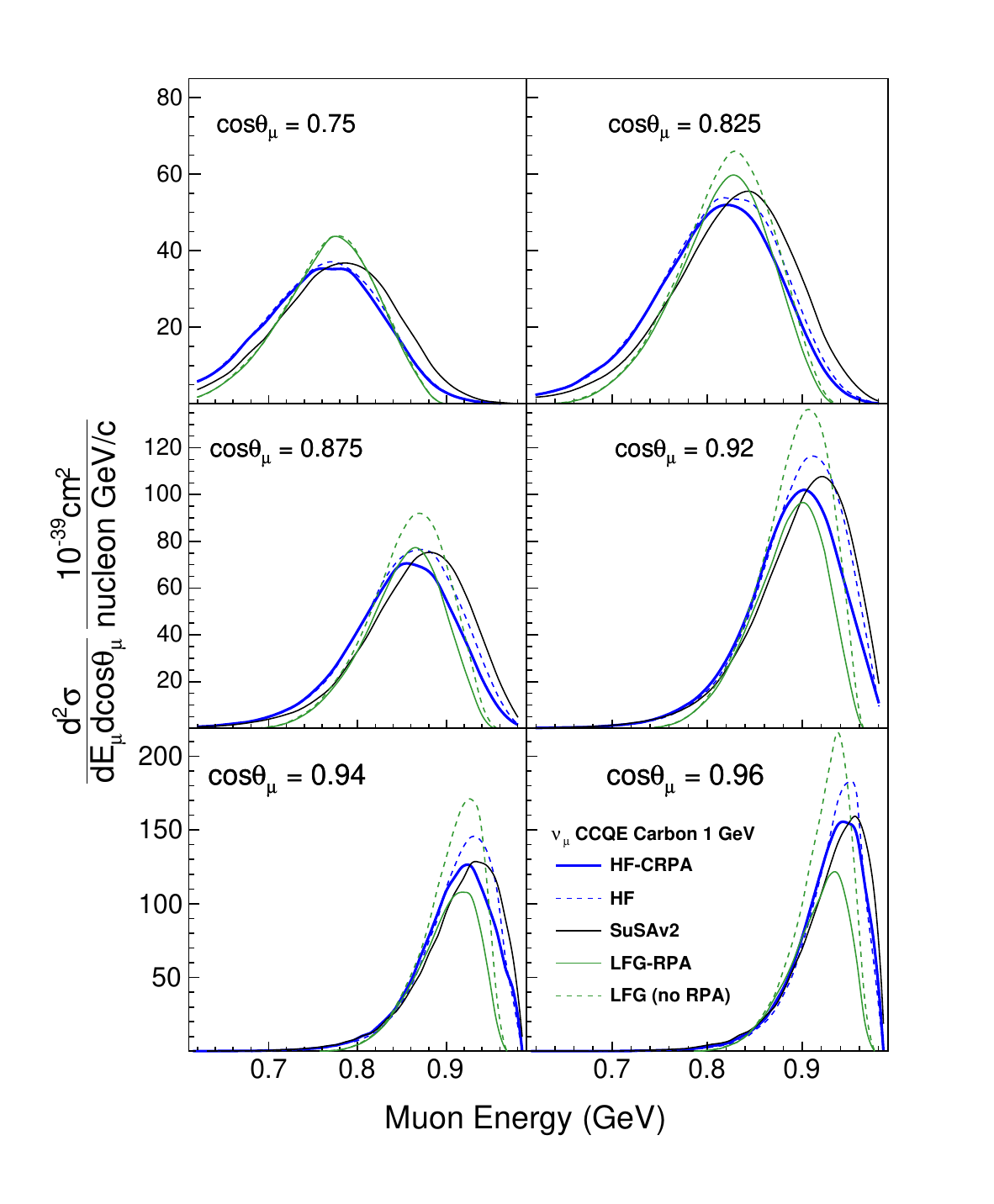}
\includegraphics[width=0.48\textwidth]{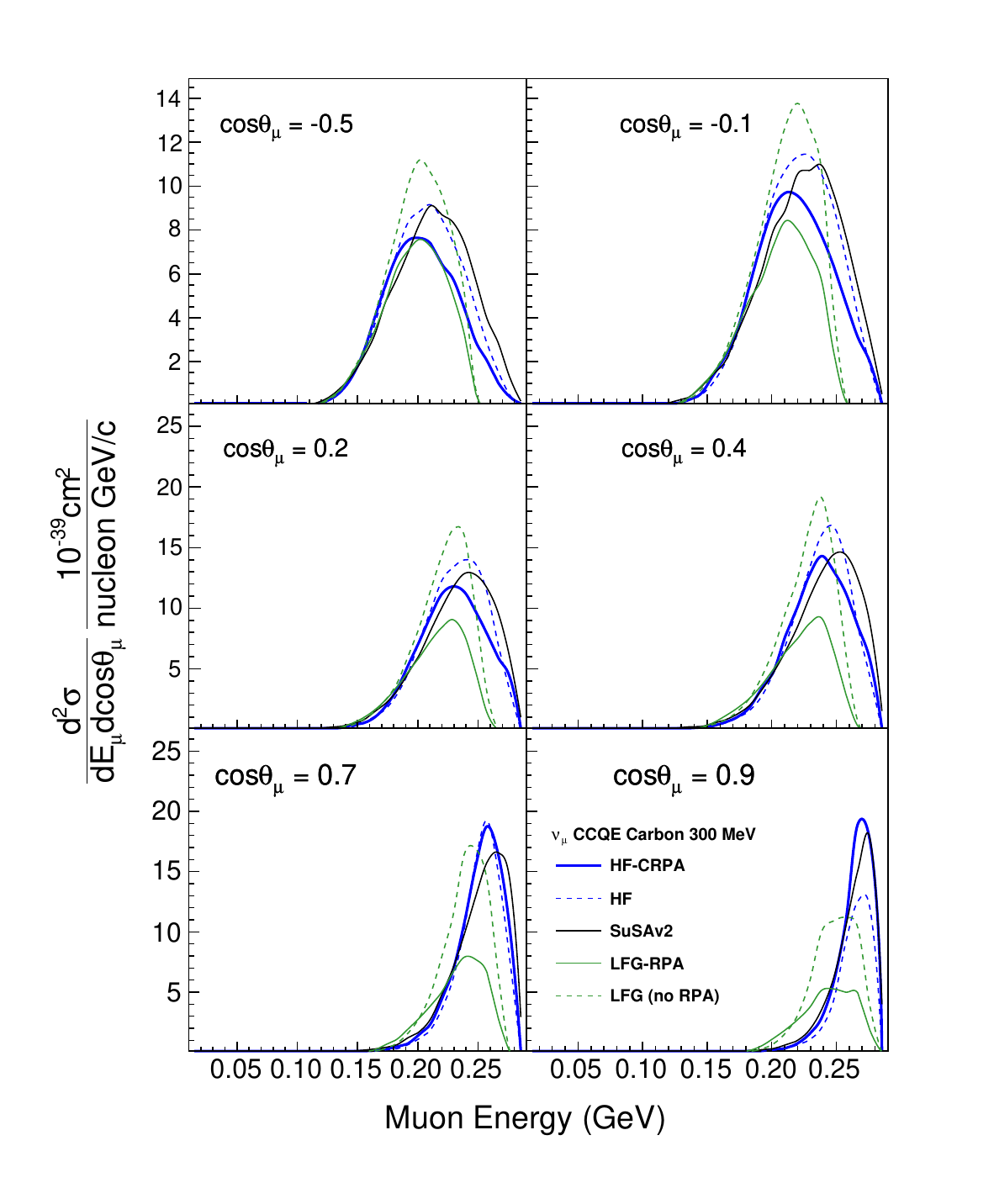}\vspace{-10mm}
\end{center}
\caption{A comparison of the muon neutrino CCQE double differential cross section on carbon as a function of outgoing muon kinematics predicted by various models with an incoming neutrino energy of either 1 GeV (left) or 300 MeV (right).}
\label{fig:monoEFlavourComp}
\end{figure*}

Fig.~\ref{fig:monoEFlavourComp} shows the comparison of the double differential charged-current $\nu_\mu$ scattering cross section obtained with several models at fixed incoming energies. RPA suppression in both the LFG-RPA and HF-CRPA models are largest at forward scattering angles and decreases for backward angles. RPA effects are most important at small incoming neutrino energies, but still affect the cross section non trivially for larger energies. One sees that the RPA in the Valencia model is particularly strong, and leads mostly to a suppression of the cross section within the angular ranges considered. In the HF-CRPA model on the other hand, the RPA leads to a milder suppression, and a shift of the cross section towards larger excitation energies as demonstrated by the shift of the peak to lower muon energies in HF-CRPA compared to HF. 

\begin{figure*}[!tb]
\begin{center}
\includegraphics[width=0.48\textwidth]{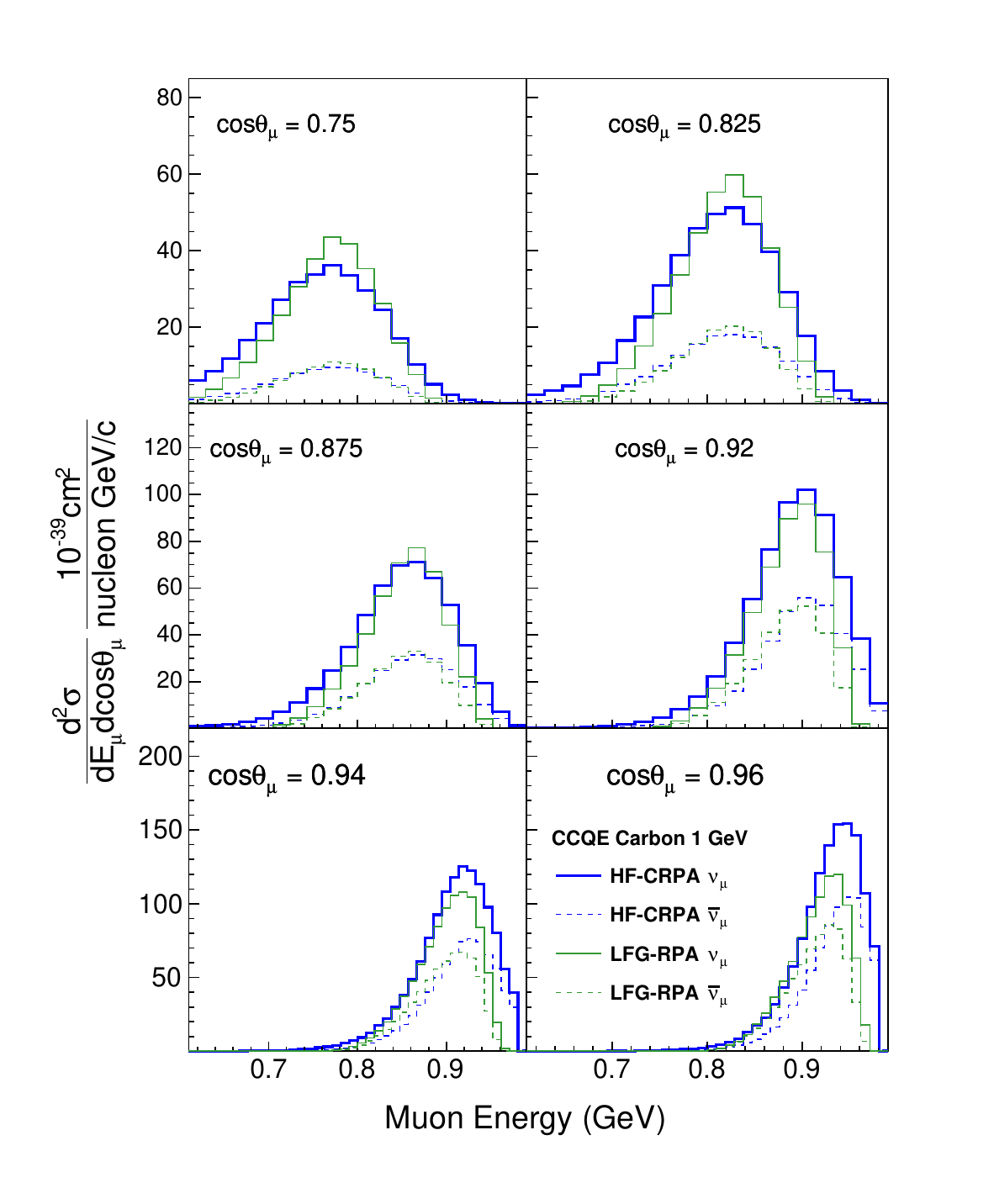}
\includegraphics[width=0.48\textwidth]{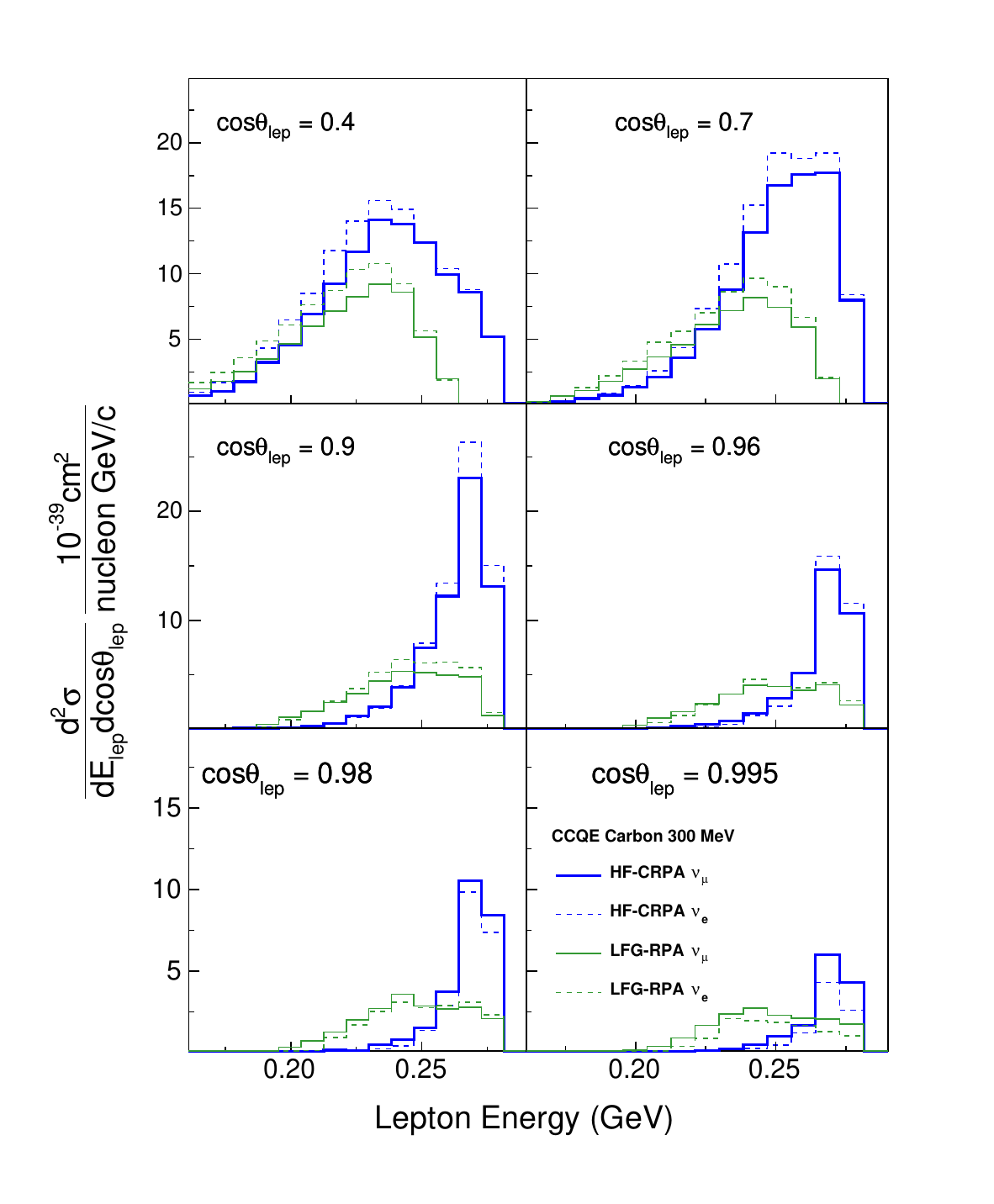}\vspace{-10mm}
\end{center}
\caption{A comparison of CCQE double differential cross section on carbon as a function of outgoing muon kinematics predicted by various models. The left plot compares the neutrino and anti-neutrino cross sections predicted for an incoming neutrino energy of 1 GeV, whilst the right plot shows a comparison between the predicted muon and electron neutrino cross sections for an incoming neutrino energy of 300 MeV. The right plot is drawn as a binned histogram for an easier comparison due to tightly overlapping curves and the available event generator statistics. Note that the scattering angles considered in the two cases are different.}
\label{fig:monoEComp}
\end{figure*}

Fig.~\ref{fig:monoEComp} additionally highlights other important differences between the Valencia LFG-RPA and HF-CRPA models. The left panels compare the neutrino and anti-neutrino cross sections for 1 GeV incoming neutrinos. Whilst these are generally similar at low energy transfer, intermediate muon energies show wide regions of the phase space where the neutrino cross section for HF-CRPA is larger than that of LFG-RPA, whilst this is inverted for the anti-neutrino case.  The right panels of Fig.~\ref{fig:monoEComp} show a comparison of the muon and electron neutrino cross sections for 300 MeV incoming neutrinos. It can immediately be noted that the strong aforementioned LFG-RPA suppression causes a large difference in the model predictions, but it can also be seen that the ratio of the muon and electron neutrino cross sections is quite different at forward angles. As described in Sec.~\ref{sec:intro}, such differences could have implications for future high statistics analyses of CPV. The quantification of the such effects will be studied in future work.

\subsection{Analysis of T2K measurements}\label{sec:t2kcomp}

As described in Sec.~\ref{sec:intro}, each model is extended from CCQE-only to CC0$\pi$, by adding  SuSAv2 2p2h and standard GENIE pion absorption components, such that they can be compared to model independent experimental data (note that a very small CCQE contribution is also removed due to pion production FSI). In view of exploring model differences between different nuclear targets and flavours, the models are compared to T2K measurements of the CC0$\pi$ cross section made simultaneously for either carbon and oxygen targets~\cite{T2K:2020jav} or for neutrinos and anti-neutrinos~\cite{T2K:2020sbd} using the NUISANCE framework~\cite{Stowell:2016jfr}.  The ability for each model to describe the data is shown as a $\chi^2$ score calculated using the full experimental covariance matrix in Tab.~\ref{tab:chi2}. Note that all the plots shown do not include all experimental bins. Very high momentum bins with large uncertainties in both measurements are not shown and for the neutrino and anti-neutrino case in Sec.~\ref{sec:nuandanu} some of the high angle slices are also not shown since the focus of the discussion concerns the more forward angular region. The $\chi^2$ score includes all bins.

\begin{center} 
\begin{table}[!htbp]
\begin{tabular}{ |l|c|c| } 
 \hline
  & Carbon and oxygen & $\nu_\mu$ and $\bar{\nu}_\mu$  \\
 \hline
Number of bins & 58 & 116 \\
 \hline
HF-CRPA & 135 & 740 \\
HF & 143 & 683 \\
SuSAv2 & 140 & 741 \\
LFG-RPA & 59 & 446 \\
LFG (no RPA) & 184 & 1028 \\
\hline
\end{tabular}
\caption{The $\chi^2$ calculated from comparing each model to T2K $\nu_\mu$ CC0$\pi$ cross section measurements jointly on carbon and oxygen targets~\cite{T2K:2020jav} or for neutrinos and anti-neutrinos~\cite{T2K:2020sbd} interactions on hydrocarbon. Note that the SuSAv2 2p2h and GENIE pion absorption contributions are added to each model for comparison with the data.}
 \label{tab:chi2}
\end{table}
\end{center}

\subsubsection{Carbon and Oxygen}\label{sec:cando}

\begin{figure}[htbp]
\begin{center}
\includegraphics[width=0.48\textwidth]{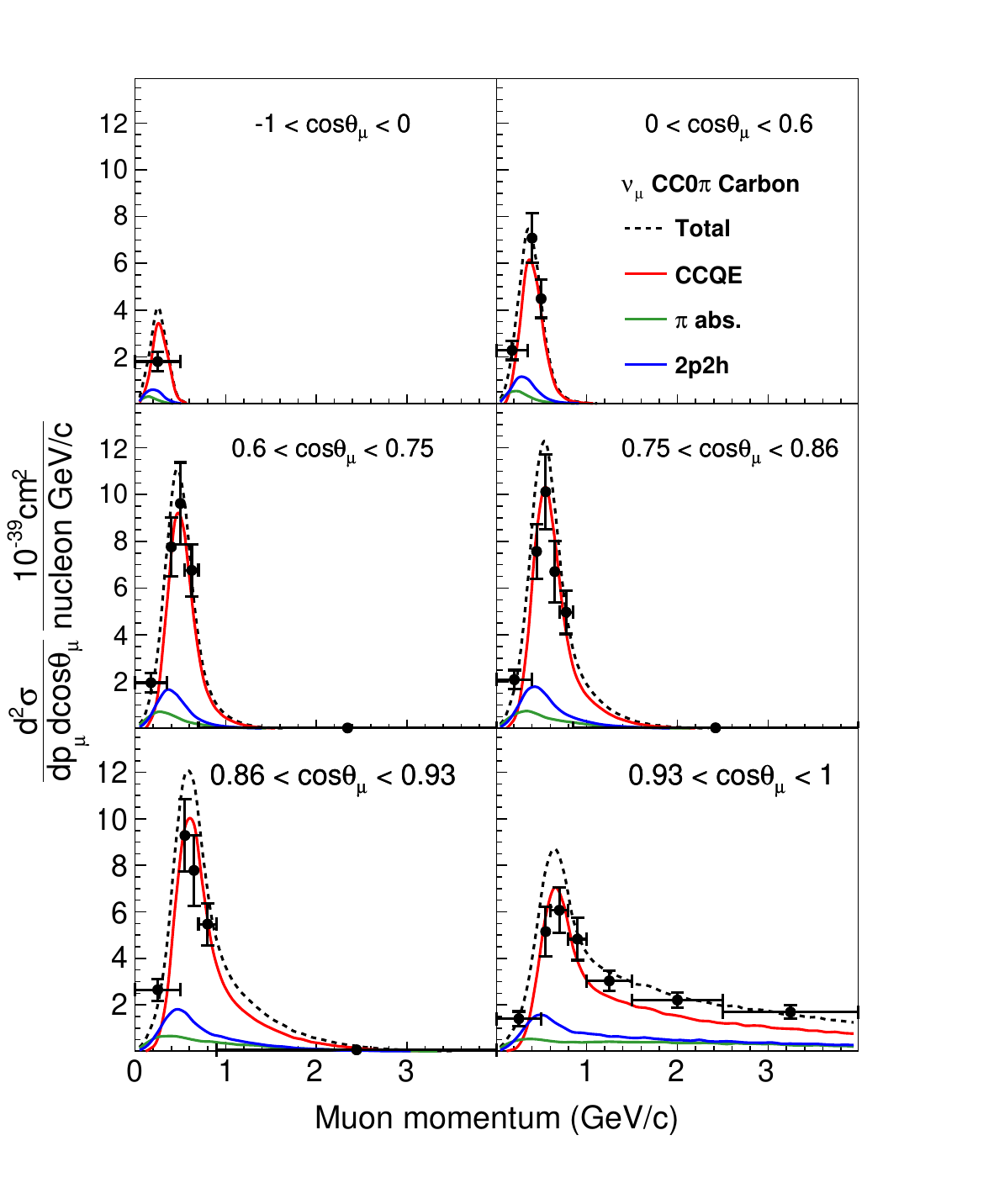}
\end{center}
\vspace{-10mm}
\caption{The T2K flux-integrated CC0$\pi$ $\nu_\mu$ double differential cross section on carbon as a function of outgoing muon kinematics as predicted by the newly implemented HF-CRPA model compared to T2K measurements~\cite{T2K:2020jav}. The prediction is split by interaction mode. Note that the new implementation is only the CCQE contribution, the non-CCQE contributions are described in Sec~\ref{sec:intro}.}
\label{fig:T2K_C_byMode}
\end{figure}

\begin{figure*}[htbp]
\begin{center}
\includegraphics[width=0.48\textwidth]{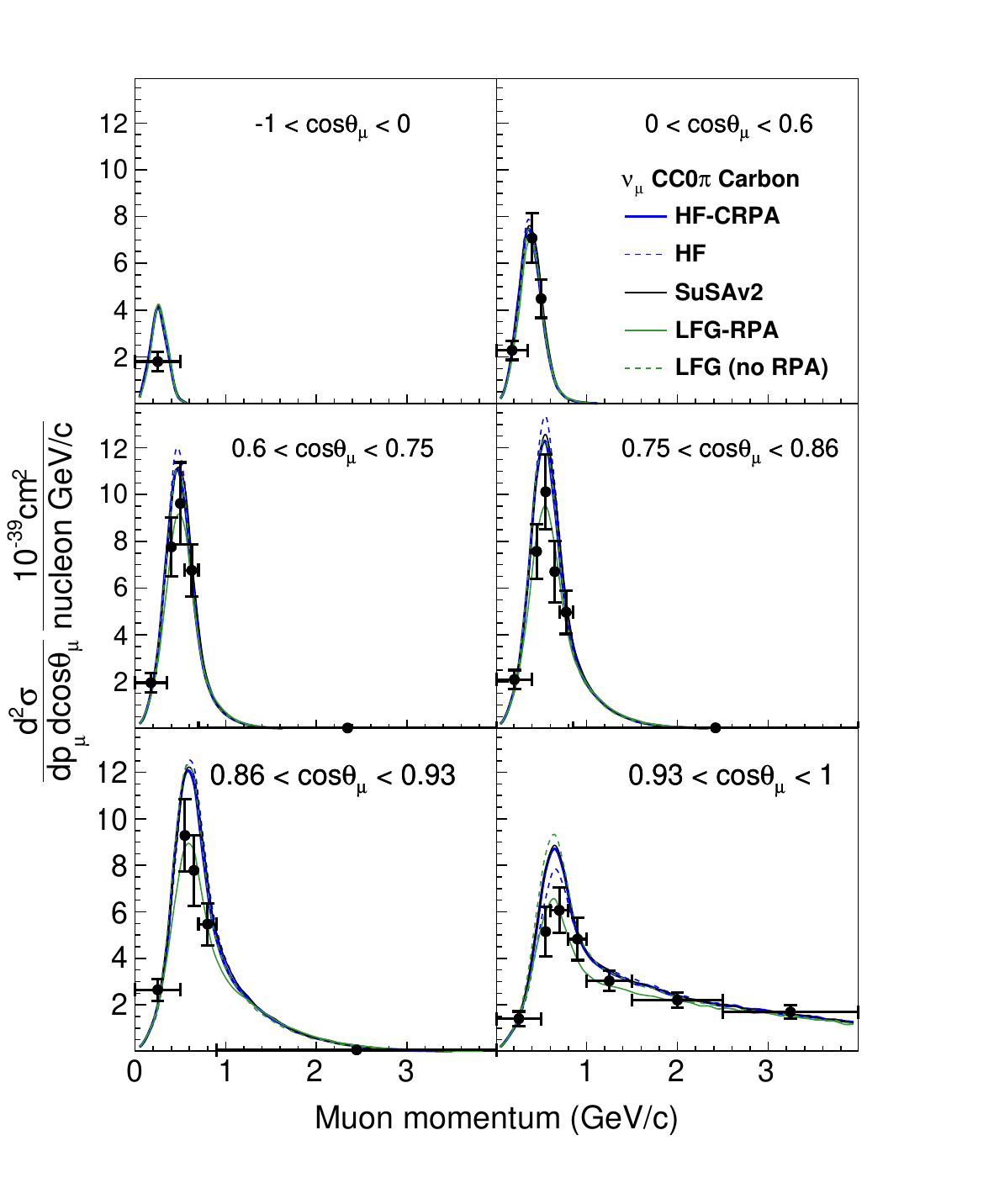}
\includegraphics[width=0.48\textwidth]{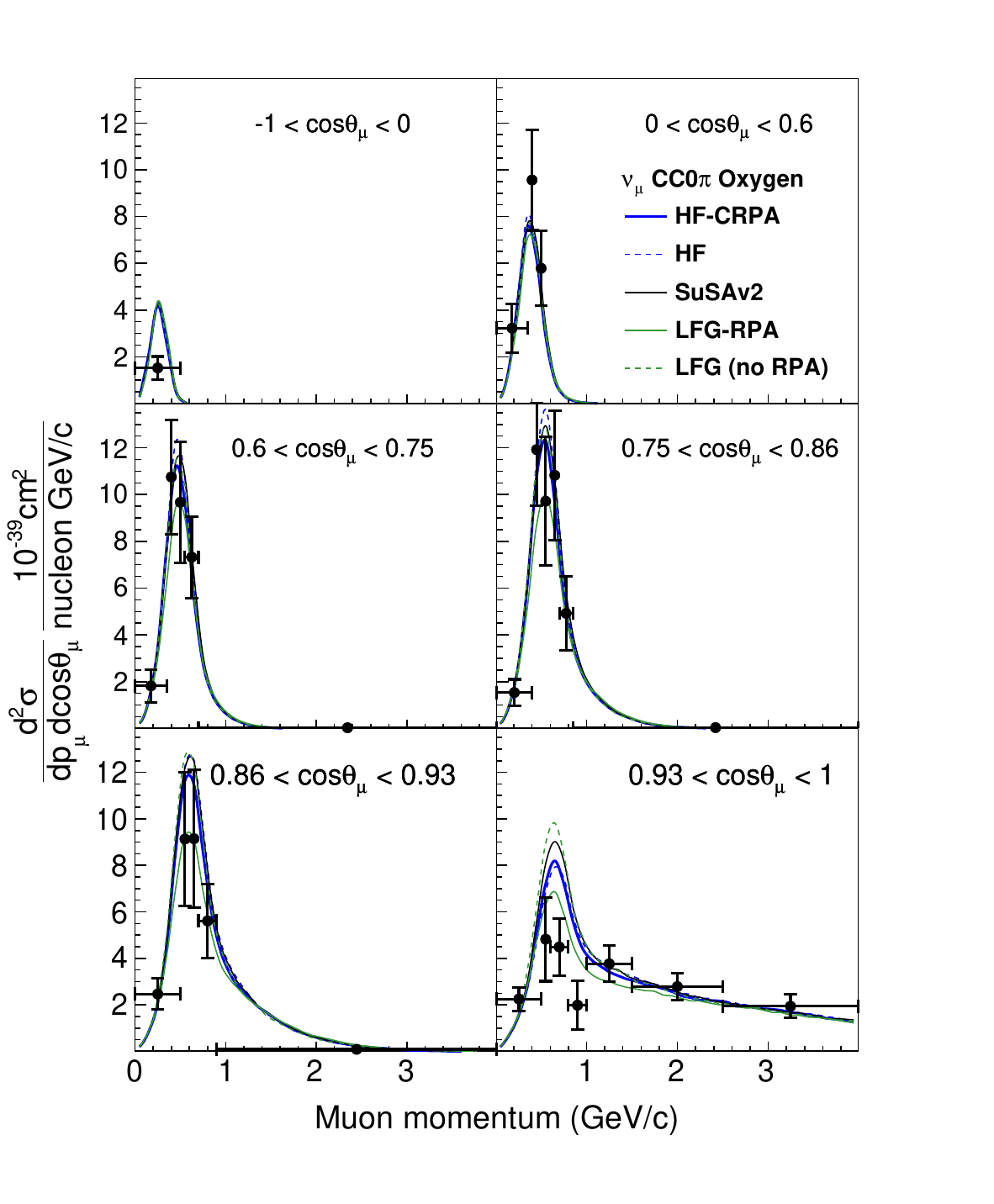}\vspace{-10mm}
\end{center}
\caption{The T2K flux-integrated CC0$\pi$ $\nu_\mu$ double differential cross section on carbon and oxygen as a function of outgoing muon kinematics predicted by various models compared to T2K measurements~\cite{T2K:2020jav}. Note that the new implementation is only the CCQE contribution, the non-CCQE contributions are described in Sec~\ref{sec:intro}.}
\label{fig:T2K_CandO_modelComp}
\end{figure*}

Fig.~\ref{fig:T2K_C_byMode} shows the T2K $\nu_\mu$ CC0$\pi$ cross-section measurement on carbon compared to HF-CRPA predictions, including the additional 2p2h and pion absorption contributions, split by interaction mode whilst Fig.~\ref{fig:T2K_CandO_modelComp} shows a comparison of all the considered GENIE models to the full carbon and oxygen measurement. 

It can be noted that, at large angles ($\cos\theta_\mu$$<$0.6) which corresponds to regions of large $\omega$ and $q$, all the models are in excellent agreement with each other and reasonable agreement with the T2K measurement. In this region the cross section is driven mostly by nucleon-level physics, related to the choice of form factors which are very similar for all the considered models (all use a dipole axial form factor with a nucleon axial mass close to 1~GeV). At more forward angles (and therefore correspondingly lower $\omega$, $q$) nuclear effects become more important and the models begin to differ. As discussed in the context of Fig.~\ref{fig:q0q3}, it's clear how RPA has a large impact even at intermediate angles (0.6$<$$\cos\theta_\mu$$<$0.86). The largest model differences are seen in the very forward region, where the treatment of FSI effects in HF/HF-CRPA and deviations from the impulse approximation are most important. In the most forward regions, no model can describe the data for the oxygen cross section whilst only the strong suppression from RPA can describe the carbon results. However, it is clear from Fig.~\ref{fig:T2K_C_byMode} that the poor agreement between HF-CRPA and the data may be due to the mismodelling of interaction modes beyond CCQE. New exclusive analyses, such as those presented by the MINERvA collaboration in Ref.~\cite{MINERvA:2022mnw}, may be able to use outgoing nucleon kinematics to determine whether the over-prediction of the data at forward angles is concentrated at kinematics best matching CCQE or other interaction channels.  

\begin{figure*}[htbp]
\begin{center}
\includegraphics[width=0.48\textwidth]{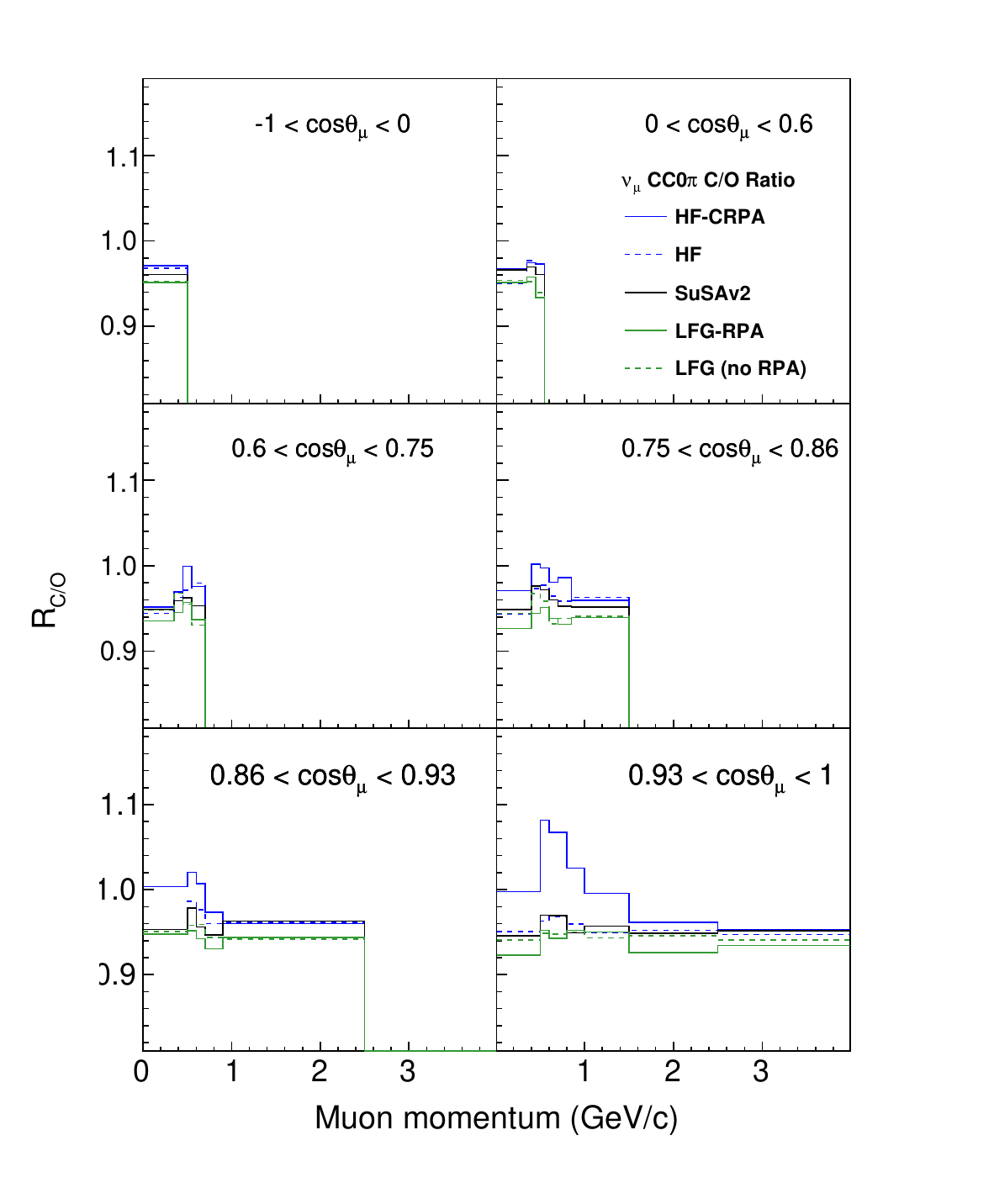}
\includegraphics[width=0.48\textwidth]{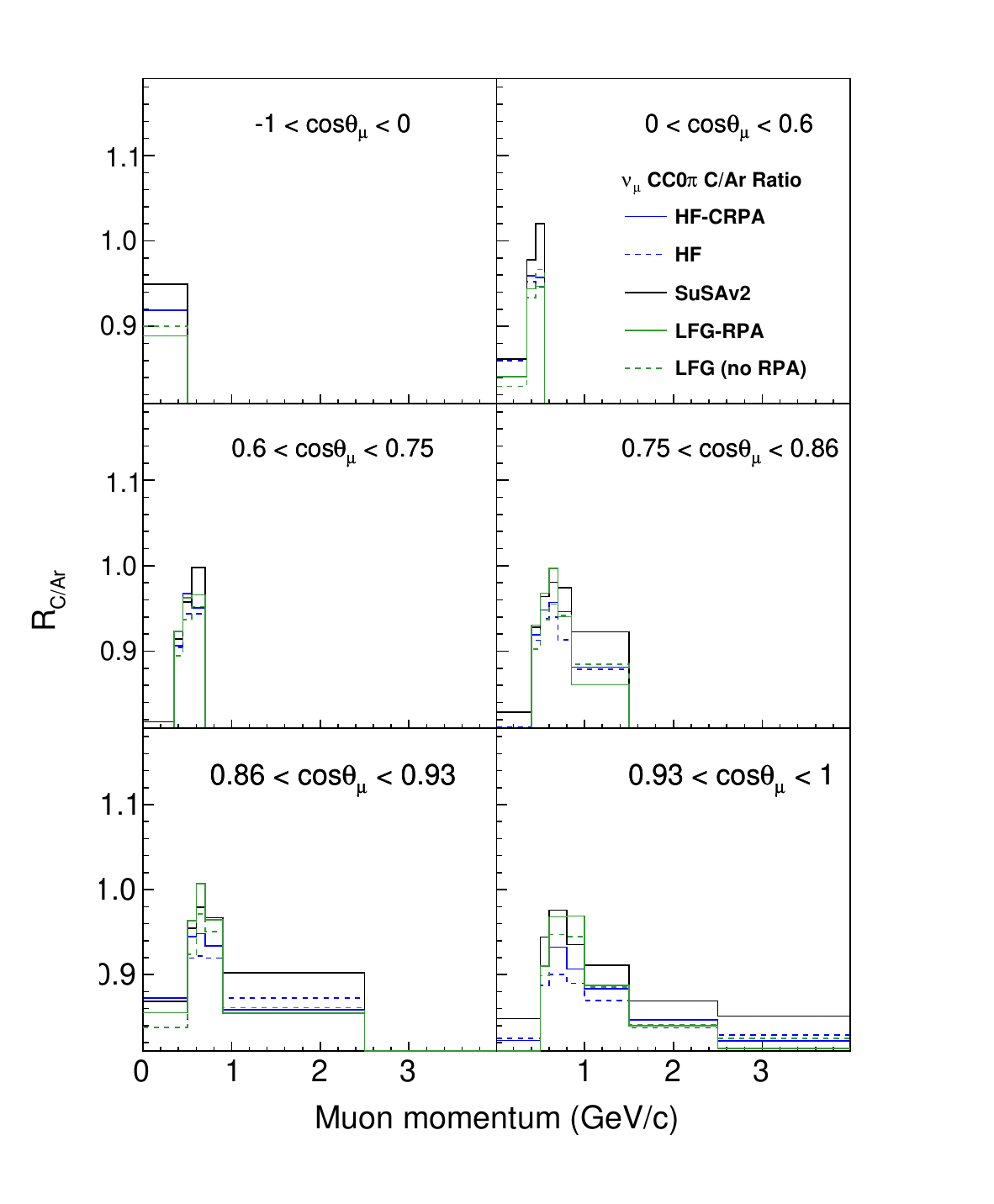}\vspace{-10mm}
\end{center}
\caption{The T2K flux-integrated CC0$\pi$ $\nu_\mu$ double differential cross section ratio between carbon and oxygen (left) or argon (right) as predicted by various models. Bins in which the cross section is too small for a meaningful ratio to be calculated are left empty. Note that the new implementation is only the CCQE contribution, the non-CCQE contributions are described in Sec~\ref{sec:intro}, and that the cross sections prior to the ratio were calculated per nucleon (rather than per neutron).}
\label{fig:T2K_CandO_ratio}
\end{figure*}

Fig.~\ref{fig:T2K_CandO_ratio} shows the prediction of the carbon/oxygen and carbon/argon cross-section ratios predicted by each model, demonstrating substantial differences. It can be noted that HF/HF-CRPA predicts that the cross section for oxygen may be lower compared to that of carbon at forward angles (and so at low $\omega$, $q$), as hinted by the T2K measurement (but with large uncertainties), whilst the ratios from the other models remain almost flat. 
This oxygen-carbon difference was previously shown for cross sections at fixed incoming energy in Ref.~\cite{VanDessel:2017ery}. Since oxygen is a double magic nucleus, a lower cross section at forward angles can be expected. As carbon and oxygen tend to have quite similar initial-state properties (e.g. binding energies and momentum distributions), this effect might be obscured in a factorised PWIA approach while it is present with a consistent treatment of initial- and final-state wavefunctions.
For the carbon to argon ratio an even larger difference is seen between the models, where it appears that CRPA has a large impact on the cross section ratio, especially at forward angles, but that RPA effects do not.

\subsubsection{Neutrino and anti-neutrino}\label{sec:nuandanu}

\begin{figure}[htbp]
\begin{center}
\includegraphics[width=0.48\textwidth]{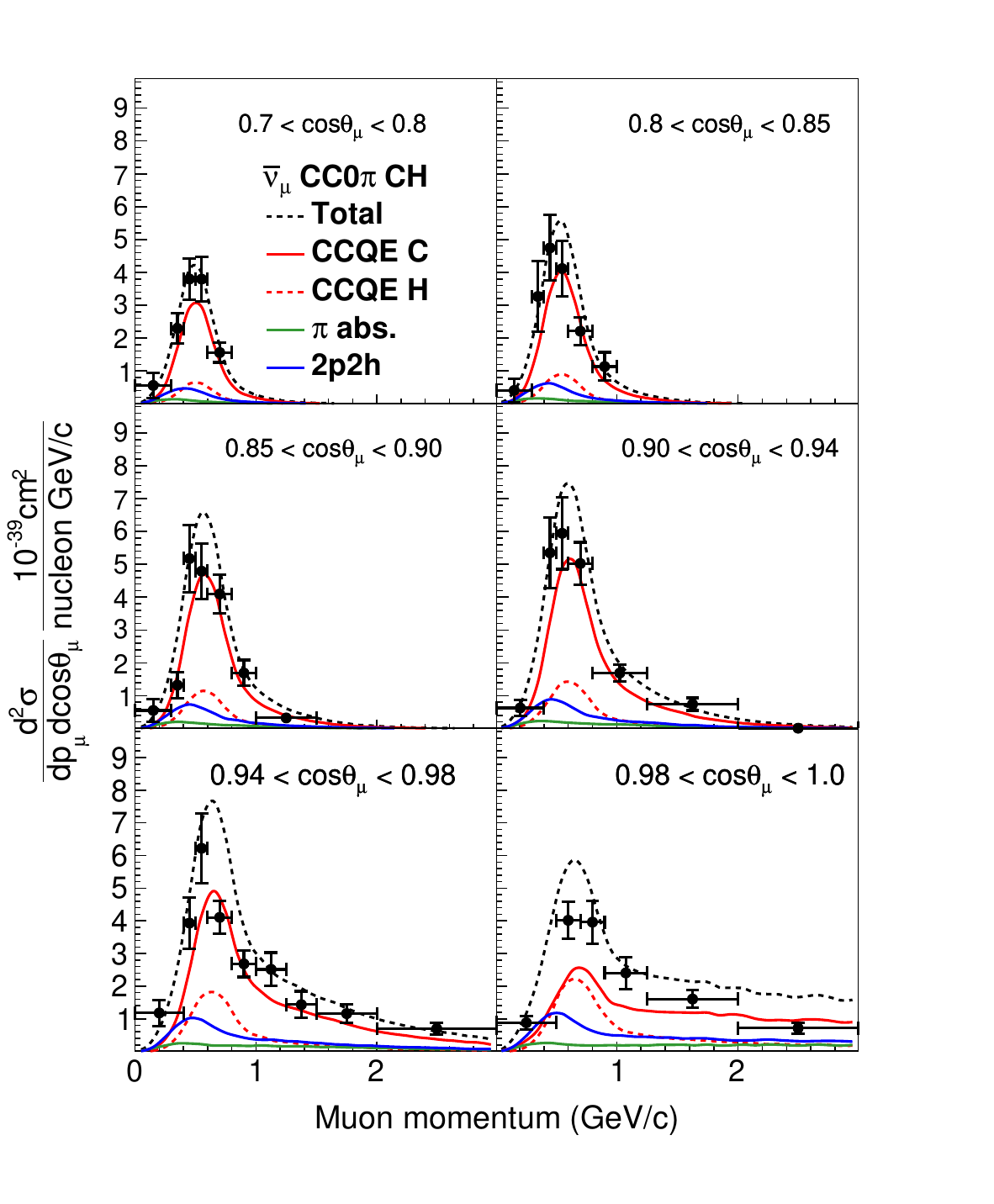} \vspace{-12mm}
\end{center}
\caption{The T2K flux-integrated CC0$\pi$ $\bar{\nu}_\mu$ double differential cross section on hydrocarbon (CH) as a function of outgoing muon kinematics as predicted by the newly implemented HF-CRPA model compared to T2K measurements~\cite{T2K:2020sbd}. The prediction is split by interaction mode and, for the CCQE case, into whether the interaction is with a carbon or hydrogen target nucleus. Note that the new implementation is only the CCQE contribution, the non-CCQE contributions are described in Sec~\ref{sec:intro}.}
\label{fig:T2K_CH-anu_byMode}
\end{figure}

\begin{figure*}[htbp]
\begin{center}
\includegraphics[width=0.48\textwidth]{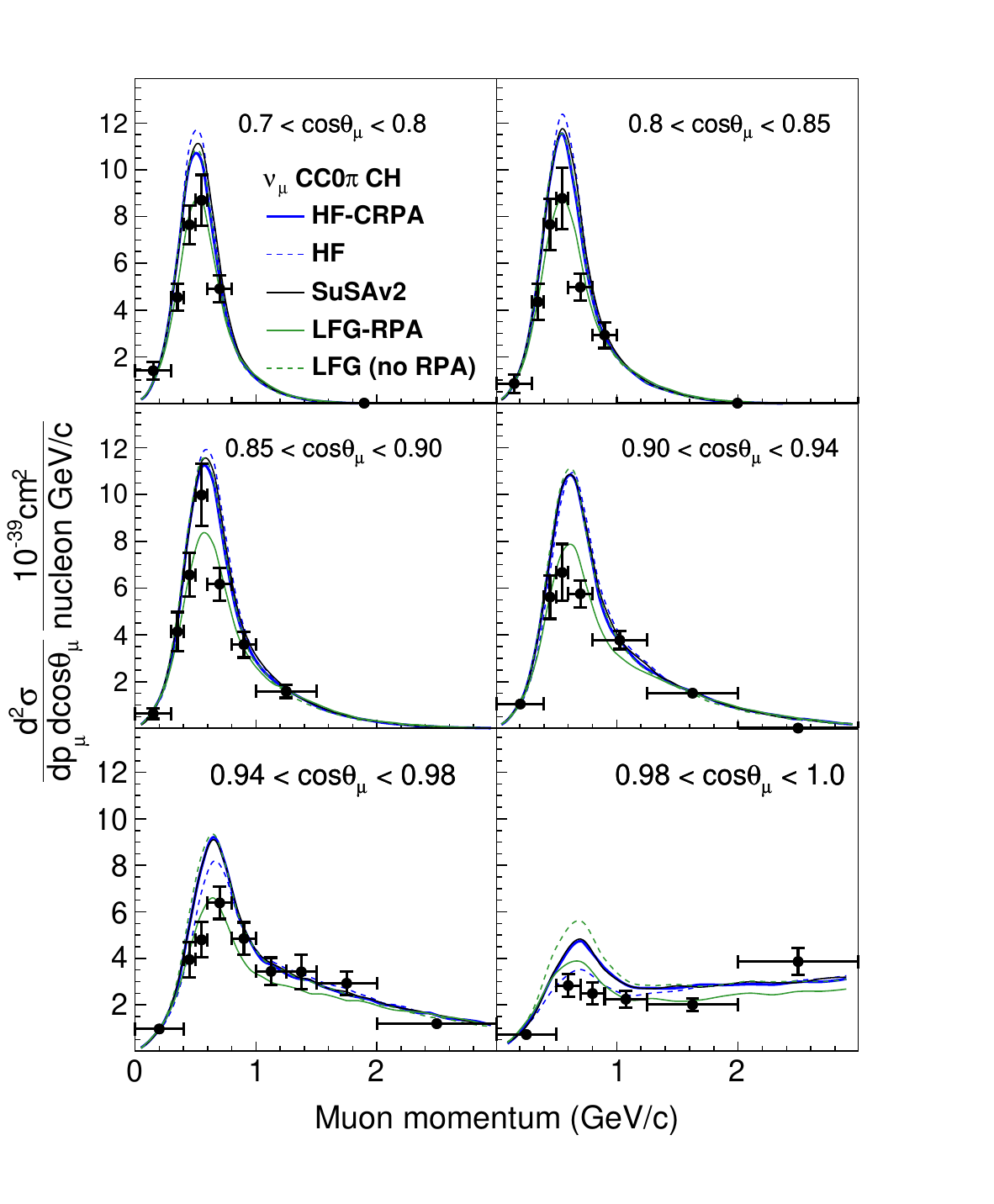}
\includegraphics[width=0.48\textwidth]{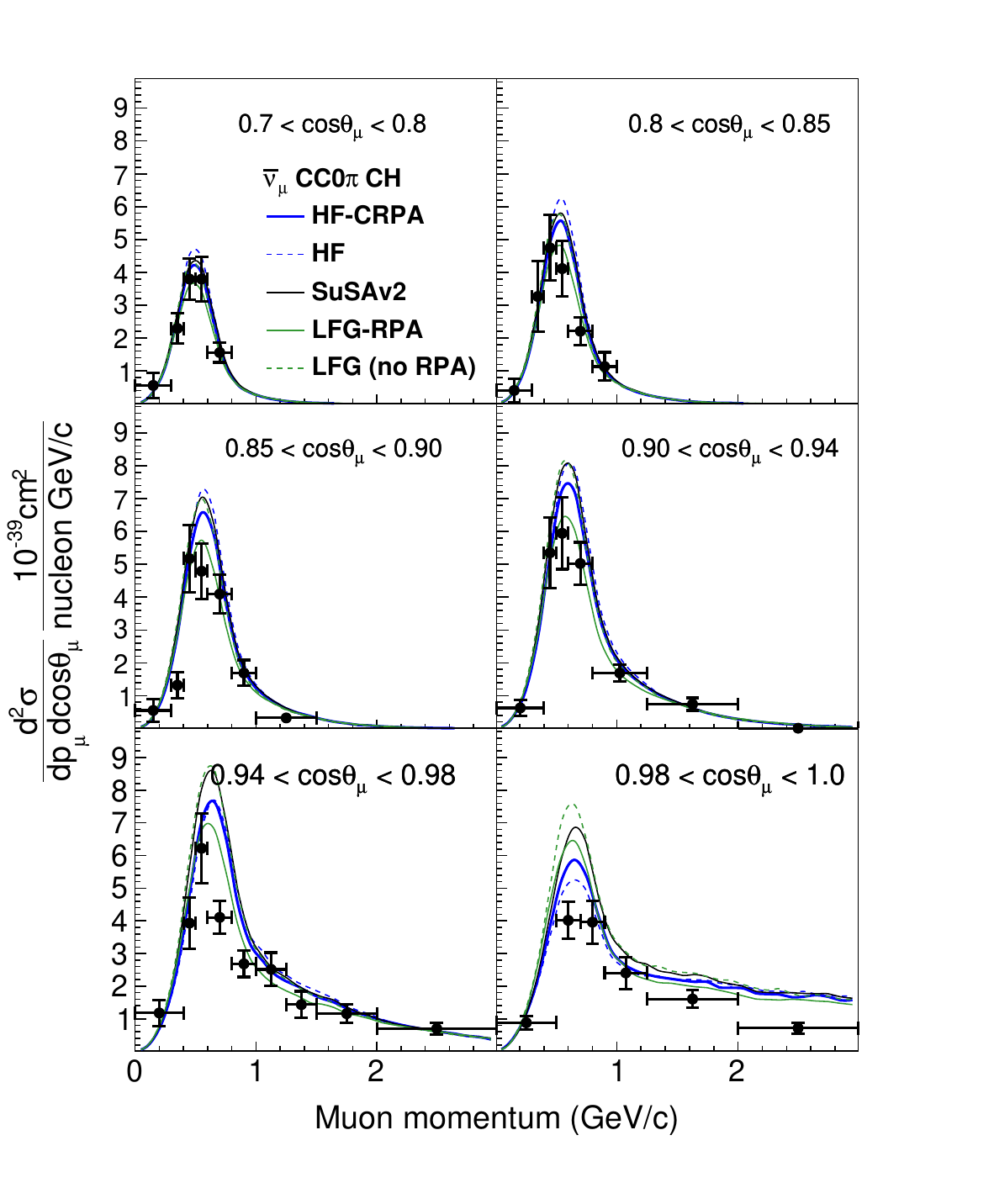}\vspace{-10mm}
\end{center}
\caption{The T2K flux-integrated CC0$\pi$ $\nu_\mu$ and $\bar{\nu}_\mu$ double differential cross section on carbon and oxygen as a function of outgoing muon kinematics predicted by various models compared to T2K measurements~\cite{T2K:2020sbd}. Note that the new implementation is only the CCQE contribution, the non-CCQE contributions are described in Sec~\ref{sec:intro}.}
\label{fig:T2K_nuandanu_modelComp}
\end{figure*}

Fig.~\ref{fig:T2K_CH-anu_byMode} shows the T2K $\bar{\nu}_\mu$ CC0$\pi$ cross-section measurement on hydrocarbon compared to HF-CRPA predictions split by interaction mode, whilst Fig.~\ref{fig:T2K_nuandanu_modelComp} shows a comparison of all the considered GENIE models to the full T2K neutrino and anti-neutrino measurement. It can be seen that for anti-neutrino interactions the hydrogen contribution (which is extremely similar between all models) is particularly significant at forward angles, where it is not subject to the same suppression from nuclear effects as the carbon contribution. As in the carbon and oxygen case, the models differ most significantly at forward angles. With the finer angular binning it is easier to note the enhancement caused by CRPA compared to the suppression caused by RPA. This is due to the most forward bins isolating both a low energy and momentum transfer where the models differ most strongly (i.e. in the lower left portion of the ratio plots shown in Fig.~\ref{fig:q0q3}). 

It can further be seen that the reduction of the cross section for anti-neutrino interactions in the most forward angular slice with respect to the penultimate slice is much stronger for HF/HF-CRPA than for LFG/LFG-RPA. It can also be seen how the impact of RPA is much stronger than CRPA for both neutrinos and anti-neutrinos. RPA corrections tend to be stronger for neutrino than for anti-neutrino, whilst CRPA shows more similar strength. 

It is clear that all models significantly overestimate the T2K anti-neutrino measurement at forward angles, with possible the exception of HF-CRPA in the most forward slice, and all models other than LFG-RPA struggle to describe the neutrino measurement. Whilst it is once again tempting to interpret this as a requirement for a strong RPA-like CCQE suppression, the non-negligible contributions from non-QE interaction modes allows an alternative resolution by a significant reduction of the pion absorption and 2p2h strengths.

\section{Discussion and conclusions}\label{sec:conc}

The demonstrated differences in model predictions for the evolution of the cross section as a function of neutrino energy, kinematics, flavour and target can imply challenges for near-to-far detector extrapolation for neutrino oscillation analyses. For example, it is clear from Fig.~\ref{fig:sigmaenu}~and~\ref{fig:q0q3} that considering nucleon correlations via the LFG-RPA or HF-CRPA approaches yields shape differences for both total and differential cross sections. This implies that constraints from a near detector may be incorrectly propagated to the far detector if nucleon correlations are mis-modelled. Similarly Fig.~\ref{fig:T2K_CandO_modelComp}~and~\ref{fig:T2K_CandO_ratio} shows that the extrapolation of cross sections from one target to another is quite dependent on the model used, especially in regions of low $\omega$, $q$. Taking the spread of the model predictions as a minimum gauge of current uncertainty would suggest that the cross-section ratio between different nuclear targets may not be controlled at better than the 5\%-10\% level. Note that this is calculated for CC0$\pi$ cross sections where the only change between models is in the QE component, and so this level of uncertainty may be greater once the impact of 2p2h and pion absorption scaling uncertainties is considered.

Fig.~\ref{fig:monoEComp}~and~\ref{fig:T2K_CH-anu_byMode} additionally demonstrate significant differences in the prediction for neutrino and anti-neutrino differences between models, also most notably at lower or intermediate $\omega$, $q$. A mismodelling of such differences can bias extrapolations of constraints from neutrino to anti-neutrino interactions from typically neutrino-dominant near detector data, potentially affecting measurements of CPV. The same argument can be made regarding the propagation of muon-neutrino measurements at a near detector to electron neutrino appearance at a far detector, which are shown to be different between models in Fig.~\ref{fig:monoEFlavourComp}.

It is clear from Tab~\ref{tab:chi2} that none of the models tested are able to describe the complete T2K measurements, typically due to overestimates of the cross section at forward angles ($\cos\theta_\mu$$>$0.8). The strong suppression from RPA seems to be favoured in the carbon and oxygen analysis although, as noted in Sec~\ref{sec:t2kcomp}, it's possible a similar reduction could be obtained by reducing the non-CCQE contributions. With this in mind, it is interesting to note the observation that the SuSAv2-MEC model for 2p2h interactions predicts a stronger contribution at forward angles compared to alternative models such as the GFMC calculation~\cite{Lovato:2020kba}, while providing a similar result for the more backward bins (see Ref.~\cite{Jachowicz:2021ieb} for a more detailed discussion).

Overall the HF-CRPA model predicts that approximately 15\% (17\%) of CCQE events in the T2K electron (muon) neutrino flux after oscillations are within the challenging low $\omega$, $q$ region where model differences are strongest (taking it to be broadly characterised by $q_3<$300 MeV/c, $q_0<$50 MeV). It is therefore clear that as experiments gather more statistics, an accurate modelling of this region is required, alongside a cautious assignment of associated systematic uncertainties. 

In conclusion, the HF and HF-CRPA models have been implemented in GENIE and give substantially different predictions from other available models, particularly at low momentum and energy transfer. It has further been shown that the predicted evolution of the cross section as a function of neutrino energy, kinematics, flavour and target all differ between the newly implemented models and the other GENIE models considered. Since neutrino oscillation measurements typically rely on the correct modelling of at least some aspects of this evolution when extrapolating constraints from a near detector to a far detector, the addition of the HF and HF-CRPA models to GENIE provide an important means to evaluate potential systematic uncertainties within future analyses.

\section*{Acknowledgements}

The authors would like to thank the GENIE collaboration and all the HF-CRPA model authors from the Ghent University group. SD is particularly grateful for fruitful conversations with Steve Dytman and technical help from Marco Roda. VP acknowledges the support from US DOE under grant DE-SC0009824. This manuscript has been authored by Fermi Research Alliance, LLC under Contract No. DE-AC02-07CH11359 with the U.S. Department of Energy, Office of Science, Office of High Energy Physics. The project was realised thanks to the CERN summer student program, which supported OP. NJ acknowledges support from the Fund for Scientific Research Flanders (FWO-Flanders).

\appendix

\section{Implementation validations}\label{app:vali}

\begin{figure*}[htbp]
\begin{center}
\includegraphics[width=0.48\textwidth]{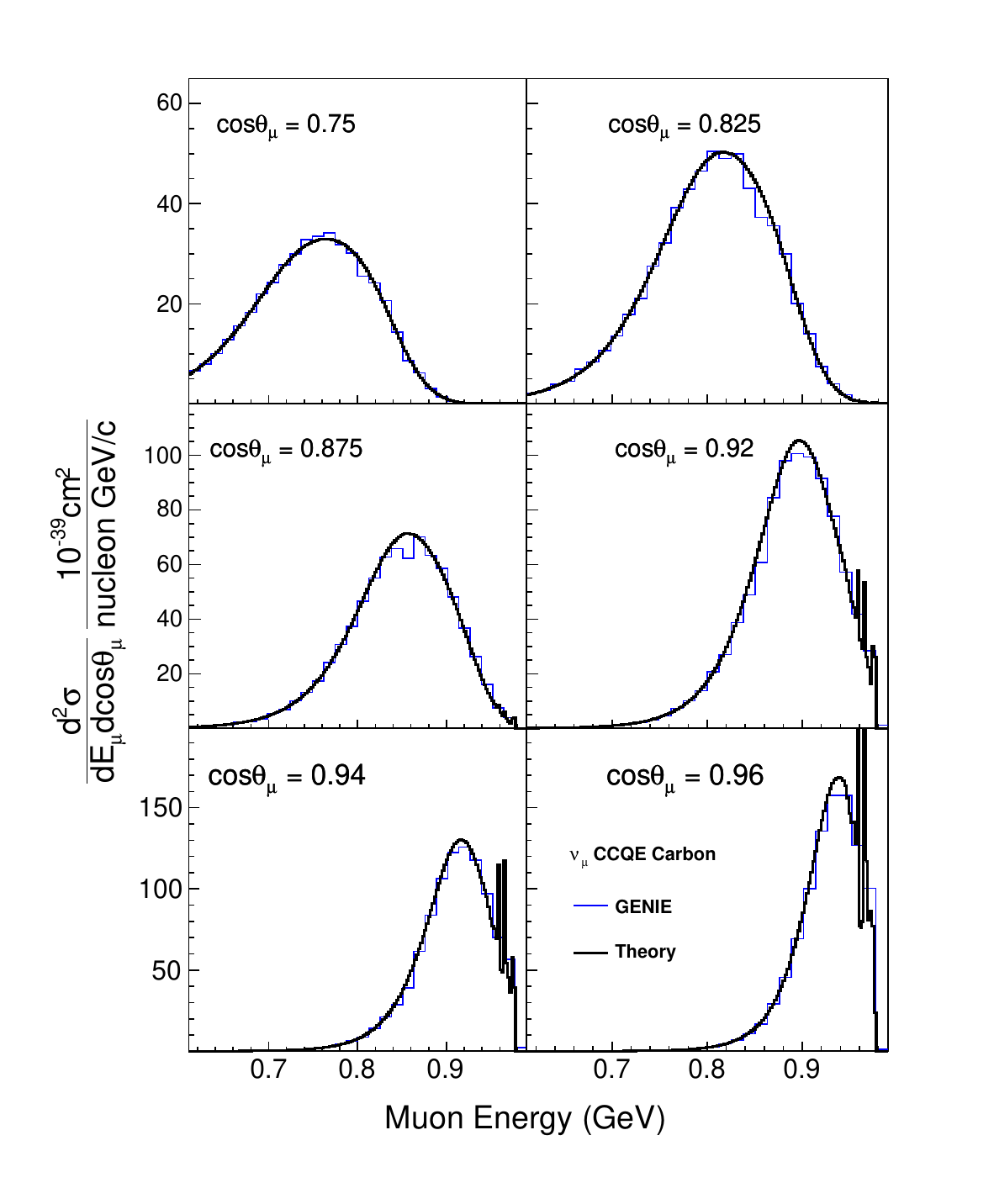}
\includegraphics[width=0.48\textwidth]{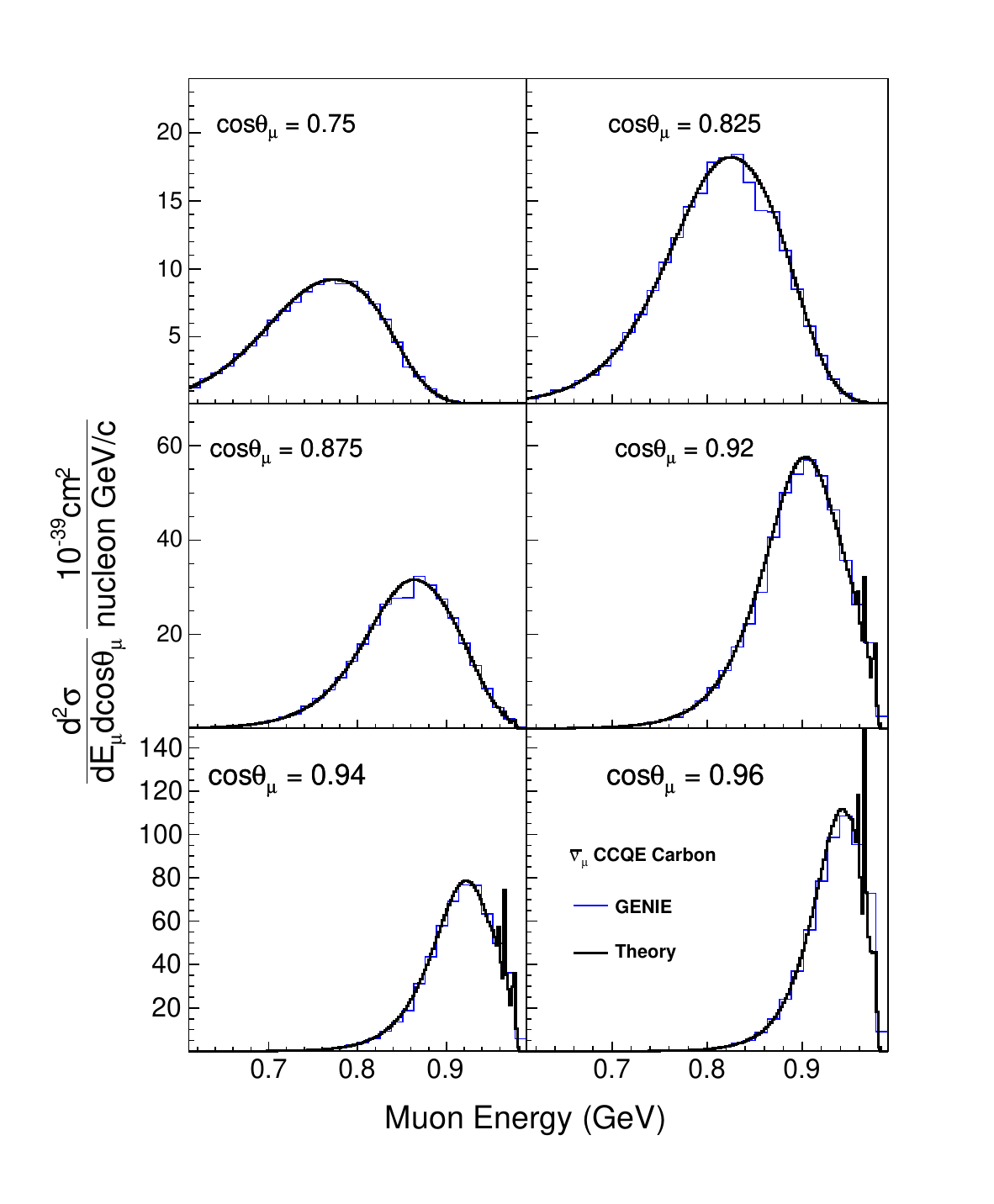}\vspace{-10mm}
\end{center}
\caption{The double differential cross section from the HF-CRPA theory code and the GENIE implementation are shown for neutrino (left) and anti-neutrino (right) CCQE interaction on a carbon target with a fixed 1 GeV incoming muon neutrino energy. }
\label{fig:vali}
\end{figure*}

The model implementation was validated to accurately reproduce both the hadron tensor elements as a function of $\omega$, $q$ and the complete double differential cross section for a variety of fixed incoming neutrino energies. An example set of validations is shown in Fig.~\ref{fig:vali}, which demonstrates a comparison of the HF-CRPA theory code and the GENIE implementation calculation of the double differential cross section on a carbon target. It can be seen that the agreement is near-perfect, with the small differences stemming from details of interpolation methods and the fact the GENIE event calculation requires a finite sized angular range in which to select events to calculate the cross section (the analysis in Fig.~\ref{fig:vali} uses a 0.02 range of $\cos\theta_\mu$). Note that the GENIE prediction, if binned fine enough, is capable of reproducing the peaks observed at forward angles within the theory predictions (at low energy transfers the energy transfer spacing of the hadron tensors is 0.25 MeV). 

\FloatBarrier

\bibliography{biblo}

\begin{thebibliography}{40}%
\makeatletter
\providecommand \@ifxundefined [1]{%
 \@ifx{#1\undefined}
}%
\providecommand \@ifnum [1]{%
 \ifnum #1\expandafter \@firstoftwo
 \else \expandafter \@secondoftwo
 \fi
}%
\providecommand \@ifx [1]{%
 \ifx #1\expandafter \@firstoftwo
 \else \expandafter \@secondoftwo
 \fi
}%
\providecommand \natexlab [1]{#1}%
\providecommand \enquote  [1]{``#1''}%
\providecommand \bibnamefont  [1]{#1}%
\providecommand \bibfnamefont [1]{#1}%
\providecommand \citenamefont [1]{#1}%
\providecommand \href@noop [0]{\@secondoftwo}%
\providecommand \href [0]{\begingroup \@sanitize@url \@href}%
\providecommand \@href[1]{\@@startlink{#1}\@@href}%
\providecommand \@@href[1]{\endgroup#1\@@endlink}%
\providecommand \@sanitize@url [0]{\catcode `\\12\catcode `\$12\catcode
  `\&12\catcode `\#12\catcode `\^12\catcode `\_12\catcode `\%12\relax}%
\providecommand \@@startlink[1]{}%
\providecommand \@@endlink[0]{}%
\providecommand \url  [0]{\begingroup\@sanitize@url \@url }%
\providecommand \@url [1]{\endgroup\@href {#1}{\urlprefix }}%
\providecommand \urlprefix  [0]{URL }%
\providecommand \Eprint [0]{\href }%
\providecommand \doibase [0]{http://dx.doi.org/}%
\providecommand \selectlanguage [0]{\@gobble}%
\providecommand \bibinfo  [0]{\@secondoftwo}%
\providecommand \bibfield  [0]{\@secondoftwo}%
\providecommand \translation [1]{[#1]}%
\providecommand \BibitemOpen [0]{}%
\providecommand \bibitemStop [0]{}%
\providecommand \bibitemNoStop [0]{.\EOS\space}%
\providecommand \EOS [0]{\spacefactor3000\relax}%
\providecommand \BibitemShut  [1]{\csname bibitem#1\endcsname}%
\let\auto@bib@innerbib\@empty
\bibitem [{\citenamefont {Abe}\ \emph {et~al.}(2011)\citenamefont {Abe} \emph
  {et~al.}}]{T2K:2011qtm}%
  \BibitemOpen
  \bibfield  {author} {\bibinfo {author} {\bibfnamefont {K.}~\bibnamefont
  {Abe}} \emph {et~al.} (\bibinfo {collaboration} {T2K}),\ }\href {\doibase
  10.1016/j.nima.2011.06.067} {\bibfield  {journal} {\bibinfo  {journal} {Nucl.
  Instrum. Meth. A}\ }\textbf {\bibinfo {volume} {659}},\ \bibinfo {pages}
  {106} (\bibinfo {year} {2011})},\ \Eprint {http://arxiv.org/abs/1106.1238}
  {arXiv:1106.1238 [physics.ins-det]} \BibitemShut {NoStop}%
\bibitem [{\citenamefont {Ayres}\ \emph {et~al.}(2007)\citenamefont {Ayres}
  \emph {et~al.}}]{NOvA:2007rmc}%
  \BibitemOpen
  \bibfield  {author} {\bibinfo {author} {\bibfnamefont {D.~S.}\ \bibnamefont
  {Ayres}} \emph {et~al.} (\bibinfo {collaboration} {NOvA}),\ }\href {\doibase
  10.2172/935497} {\  (\bibinfo {year} {2007}),\ 10.2172/935497},\ \bibinfo
  {note} {{FERMILAB-DESIGN-2007-01}}\BibitemShut {NoStop}%
\bibitem [{\citenamefont {Abe}\ \emph {et~al.}(2018)\citenamefont {Abe} \emph
  {et~al.}}]{Abe:2018uyc}%
  \BibitemOpen
  \bibfield  {author} {\bibinfo {author} {\bibfnamefont {K.}~\bibnamefont
  {Abe}} \emph {et~al.} (\bibinfo {collaboration} {Hyper-Kamiokande}),\
  }\href@noop {} {\  (\bibinfo {year} {2018})},\ \Eprint
  {http://arxiv.org/abs/1805.04163} {arXiv:1805.04163 [physics.ins-det]}
  \BibitemShut {NoStop}%
\bibitem [{\citenamefont {Abi}\ \emph {et~al.}(2020)\citenamefont {Abi} \emph
  {et~al.}}]{DUNE:2020ypp}%
  \BibitemOpen
  \bibfield  {author} {\bibinfo {author} {\bibfnamefont {B.}~\bibnamefont
  {Abi}} \emph {et~al.} (\bibinfo {collaboration} {DUNE}),\ }\href@noop {} {\
  (\bibinfo {year} {2020})},\ \Eprint {http://arxiv.org/abs/2002.03005}
  {arXiv:2002.03005 [hep-ex]} \BibitemShut {NoStop}%
\bibitem [{\citenamefont {Alvarez-Ruso}\ \emph {et~al.}(2018)\citenamefont
  {Alvarez-Ruso} \emph {et~al.}}]{Alvarez-Ruso:2017oui}%
  \BibitemOpen
  \bibfield  {author} {\bibinfo {author} {\bibfnamefont {L.}~\bibnamefont
  {Alvarez-Ruso}} \emph {et~al.},\ }\href {\doibase 10.1016/j.ppnp.2018.01.006}
  {\bibfield  {journal} {\bibinfo  {journal} {Prog. Part. Nucl. Phys.}\
  }\textbf {\bibinfo {volume} {100}},\ \bibinfo {pages} {1} (\bibinfo {year}
  {2018})},\ \Eprint {http://arxiv.org/abs/1706.03621} {arXiv:1706.03621
  [hep-ph]} \BibitemShut {NoStop}%
\bibitem [{\citenamefont {Jachowicz}\ \emph {et~al.}(2002)\citenamefont
  {Jachowicz}, \citenamefont {Heyde}, \citenamefont {Ryckebusch},\ and\
  \citenamefont {Rombouts}}]{Jachowicz:2002rr}%
  \BibitemOpen
  \bibfield  {author} {\bibinfo {author} {\bibfnamefont {N.}~\bibnamefont
  {Jachowicz}}, \bibinfo {author} {\bibfnamefont {K.}~\bibnamefont {Heyde}},
  \bibinfo {author} {\bibfnamefont {J.}~\bibnamefont {Ryckebusch}}, \ and\
  \bibinfo {author} {\bibfnamefont {S.}~\bibnamefont {Rombouts}},\ }\href
  {\doibase 10.1103/PhysRevC.65.025501} {\bibfield  {journal} {\bibinfo
  {journal} {Phys. Rev. C}\ }\textbf {\bibinfo {volume} {65}},\ \bibinfo
  {pages} {025501} (\bibinfo {year} {2002})}\BibitemShut {NoStop}%
\bibitem [{\citenamefont {Pandey}\ \emph {et~al.}(2015)\citenamefont {Pandey},
  \citenamefont {Jachowicz}, \citenamefont {Van~Cuyck}, \citenamefont
  {Ryckebusch},\ and\ \citenamefont {Martini}}]{Pandey:2014tza}%
  \BibitemOpen
  \bibfield  {author} {\bibinfo {author} {\bibfnamefont {V.}~\bibnamefont
  {Pandey}}, \bibinfo {author} {\bibfnamefont {N.}~\bibnamefont {Jachowicz}},
  \bibinfo {author} {\bibfnamefont {T.}~\bibnamefont {Van~Cuyck}}, \bibinfo
  {author} {\bibfnamefont {J.}~\bibnamefont {Ryckebusch}}, \ and\ \bibinfo
  {author} {\bibfnamefont {M.}~\bibnamefont {Martini}},\ }\href {\doibase
  10.1103/PhysRevC.92.024606} {\bibfield  {journal} {\bibinfo  {journal} {Phys.
  Rev. C}\ }\textbf {\bibinfo {volume} {92}},\ \bibinfo {pages} {024606}
  (\bibinfo {year} {2015})},\ \Eprint {http://arxiv.org/abs/1412.4624}
  {arXiv:1412.4624 [nucl-th]} \BibitemShut {NoStop}%
\bibitem [{\citenamefont {Nikolakopoulos}\ \emph {et~al.}(2018)\citenamefont
  {Nikolakopoulos}, \citenamefont {Martini}, \citenamefont {Ericson},
  \citenamefont {Van~Dessel}, \citenamefont {Gonz\'alez-Jim\'enez},\ and\
  \citenamefont {Jachowicz}}]{Nikolakopoulos:2018sbo}%
  \BibitemOpen
  \bibfield  {author} {\bibinfo {author} {\bibfnamefont {A.}~\bibnamefont
  {Nikolakopoulos}}, \bibinfo {author} {\bibfnamefont {M.}~\bibnamefont
  {Martini}}, \bibinfo {author} {\bibfnamefont {M.}~\bibnamefont {Ericson}},
  \bibinfo {author} {\bibfnamefont {N.}~\bibnamefont {Van~Dessel}}, \bibinfo
  {author} {\bibfnamefont {R.}~\bibnamefont {Gonz\'alez-Jim\'enez}}, \ and\
  \bibinfo {author} {\bibfnamefont {N.}~\bibnamefont {Jachowicz}},\ }\href
  {\doibase 10.1103/PhysRevC.98.054603} {\bibfield  {journal} {\bibinfo
  {journal} {Phys. Rev. C}\ }\textbf {\bibinfo {volume} {98}},\ \bibinfo
  {pages} {054603} (\bibinfo {year} {2018})},\ \Eprint
  {http://arxiv.org/abs/1808.07520} {arXiv:1808.07520 [nucl-th]} \BibitemShut
  {NoStop}%
\bibitem [{\citenamefont {Ryckebusch}\ \emph {et~al.}(1988)\citenamefont
  {Ryckebusch}, \citenamefont {Waroquier}, \citenamefont {Heyde}, \citenamefont
  {Moreau},\ and\ \citenamefont {Ryckbosch}}]{RYCKEBUSCH1988237}%
  \BibitemOpen
  \bibfield  {author} {\bibinfo {author} {\bibfnamefont {J.}~\bibnamefont
  {Ryckebusch}}, \bibinfo {author} {\bibfnamefont {M.}~\bibnamefont
  {Waroquier}}, \bibinfo {author} {\bibfnamefont {K.}~\bibnamefont {Heyde}},
  \bibinfo {author} {\bibfnamefont {J.}~\bibnamefont {Moreau}}, \ and\ \bibinfo
  {author} {\bibfnamefont {D.}~\bibnamefont {Ryckbosch}},\ }\href {\doibase
  https://doi.org/10.1016/0375-9474(88)90483-6} {\bibfield  {journal} {\bibinfo
   {journal} {Nuclear Physics A}\ }\textbf {\bibinfo {volume} {476}},\ \bibinfo
  {pages} {237} (\bibinfo {year} {1988})}\BibitemShut {NoStop}%
\bibitem [{\citenamefont {Ryckebusch}\ \emph {et~al.}(1989)\citenamefont
  {Ryckebusch}, \citenamefont {Heyde}, \citenamefont {{Van Neck}},\ and\
  \citenamefont {Waroquier}}]{RYCKEBUSCH1989694}%
  \BibitemOpen
  \bibfield  {author} {\bibinfo {author} {\bibfnamefont {J.}~\bibnamefont
  {Ryckebusch}}, \bibinfo {author} {\bibfnamefont {K.}~\bibnamefont {Heyde}},
  \bibinfo {author} {\bibfnamefont {D.}~\bibnamefont {{Van Neck}}}, \ and\
  \bibinfo {author} {\bibfnamefont {M.}~\bibnamefont {Waroquier}},\ }\href
  {\doibase https://doi.org/10.1016/0375-9474(89)90436-3} {\bibfield  {journal}
  {\bibinfo  {journal} {Nuclear Physics A}\ }\textbf {\bibinfo {volume}
  {503}},\ \bibinfo {pages} {694} (\bibinfo {year} {1989})}\BibitemShut
  {NoStop}%
\bibitem [{\citenamefont {Jachowicz}\ and\ \citenamefont
  {Nikolakopoulos}(2021)}]{Jachowicz:2021ieb}%
  \BibitemOpen
  \bibfield  {author} {\bibinfo {author} {\bibfnamefont {N.}~\bibnamefont
  {Jachowicz}}\ and\ \bibinfo {author} {\bibfnamefont {A.}~\bibnamefont
  {Nikolakopoulos}},\ }\href {\doibase 10.1140/epjs/s11734-021-00286-8}
  {\bibfield  {journal} {\bibinfo  {journal} {Eur. Phys. J. ST}\ }\textbf
  {\bibinfo {volume} {230}},\ \bibinfo {pages} {4339} (\bibinfo {year}
  {2021})},\ \Eprint {http://arxiv.org/abs/2110.11321} {arXiv:2110.11321
  [nucl-th]} \BibitemShut {NoStop}%
\bibitem [{\citenamefont {Lovato}\ \emph {et~al.}(2020)\citenamefont {Lovato},
  \citenamefont {Carlson}, \citenamefont {Gandolfi}, \citenamefont {Rocco},\
  and\ \citenamefont {Schiavilla}}]{Lovato:2020kba}%
  \BibitemOpen
  \bibfield  {author} {\bibinfo {author} {\bibfnamefont {A.}~\bibnamefont
  {Lovato}}, \bibinfo {author} {\bibfnamefont {J.}~\bibnamefont {Carlson}},
  \bibinfo {author} {\bibfnamefont {S.}~\bibnamefont {Gandolfi}}, \bibinfo
  {author} {\bibfnamefont {N.}~\bibnamefont {Rocco}}, \ and\ \bibinfo {author}
  {\bibfnamefont {R.}~\bibnamefont {Schiavilla}},\ }\href {\doibase
  10.1103/PhysRevX.10.031068} {\bibfield  {journal} {\bibinfo  {journal} {Phys.
  Rev. X}\ }\textbf {\bibinfo {volume} {10}},\ \bibinfo {pages} {031068}
  (\bibinfo {year} {2020})},\ \Eprint {http://arxiv.org/abs/2003.07710}
  {arXiv:2003.07710 [nucl-th]} \BibitemShut {NoStop}%
\bibitem [{\citenamefont {Nikolakopoulos}\ \emph {et~al.}(2019)\citenamefont
  {Nikolakopoulos}, \citenamefont {Jachowicz}, \citenamefont {Van~Dessel},
  \citenamefont {Niewczas}, \citenamefont {Gonz\'alez-Jim\'enez}, \citenamefont
  {Ud\'\i{}as},\ and\ \citenamefont {Pandey}}]{Nikolakopoulos:2019qcr}%
  \BibitemOpen
  \bibfield  {author} {\bibinfo {author} {\bibfnamefont {A.}~\bibnamefont
  {Nikolakopoulos}}, \bibinfo {author} {\bibfnamefont {N.}~\bibnamefont
  {Jachowicz}}, \bibinfo {author} {\bibfnamefont {N.}~\bibnamefont
  {Van~Dessel}}, \bibinfo {author} {\bibfnamefont {K.}~\bibnamefont
  {Niewczas}}, \bibinfo {author} {\bibfnamefont {R.}~\bibnamefont
  {Gonz\'alez-Jim\'enez}}, \bibinfo {author} {\bibfnamefont {J.~M.}\
  \bibnamefont {Ud\'\i{}as}}, \ and\ \bibinfo {author} {\bibfnamefont
  {V.}~\bibnamefont {Pandey}},\ }\href {\doibase
  10.1103/PhysRevLett.123.052501} {\bibfield  {journal} {\bibinfo  {journal}
  {Phys. Rev. Lett.}\ }\textbf {\bibinfo {volume} {123}},\ \bibinfo {pages}
  {052501} (\bibinfo {year} {2019})},\ \Eprint
  {http://arxiv.org/abs/1901.08050} {arXiv:1901.08050 [nucl-th]} \BibitemShut
  {NoStop}%
\bibitem [{\citenamefont {Andreopoulos}\ \emph {et~al.}(2010)\citenamefont
  {Andreopoulos} \emph {et~al.}}]{Andreopoulos:2009rq}%
  \BibitemOpen
  \bibfield  {author} {\bibinfo {author} {\bibfnamefont {C.}~\bibnamefont
  {Andreopoulos}} \emph {et~al.},\ }\href {\doibase 10.1016/j.nima.2009.12.009}
  {\bibfield  {journal} {\bibinfo  {journal} {Nucl. Instrum. Meth.}\ }\textbf
  {\bibinfo {volume} {A614}},\ \bibinfo {pages} {87} (\bibinfo {year}
  {2010})},\ \Eprint {http://arxiv.org/abs/0905.2517} {arXiv:0905.2517
  [hep-ph]} \BibitemShut {NoStop}%
\bibitem [{\citenamefont {Andreopoulos}\ \emph {et~al.}(2015)\citenamefont
  {Andreopoulos}, \citenamefont {Barry}, \citenamefont {Dytman}, \citenamefont
  {Gallagher}, \citenamefont {Golan}, \citenamefont {Hatcher}, \citenamefont
  {Perdue},\ and\ \citenamefont {Yarba}}]{Andreopoulos:2015wxa}%
  \BibitemOpen
  \bibfield  {author} {\bibinfo {author} {\bibfnamefont {C.}~\bibnamefont
  {Andreopoulos}}, \bibinfo {author} {\bibfnamefont {C.}~\bibnamefont {Barry}},
  \bibinfo {author} {\bibfnamefont {S.}~\bibnamefont {Dytman}}, \bibinfo
  {author} {\bibfnamefont {H.}~\bibnamefont {Gallagher}}, \bibinfo {author}
  {\bibfnamefont {T.}~\bibnamefont {Golan}}, \bibinfo {author} {\bibfnamefont
  {R.}~\bibnamefont {Hatcher}}, \bibinfo {author} {\bibfnamefont
  {G.}~\bibnamefont {Perdue}}, \ and\ \bibinfo {author} {\bibfnamefont
  {J.}~\bibnamefont {Yarba}},\ }\href@noop {} {\  (\bibinfo {year} {2015})},\
  \Eprint {http://arxiv.org/abs/1510.05494} {arXiv:1510.05494 [hep-ph]}
  \BibitemShut {NoStop}%
\bibitem [{\citenamefont {Gonz\'alez-Jim\'enez}\ \emph
  {et~al.}(2014)\citenamefont {Gonz\'alez-Jim\'enez}, \citenamefont {Megias},
  \citenamefont {Barbaro}, \citenamefont {Caballero},\ and\ \citenamefont
  {Donnelly}}]{Gonzalez-Jimenez:2014eqa}%
  \BibitemOpen
  \bibfield  {author} {\bibinfo {author} {\bibfnamefont {R.}~\bibnamefont
  {Gonz\'alez-Jim\'enez}}, \bibinfo {author} {\bibfnamefont {G.~D.}\
  \bibnamefont {Megias}}, \bibinfo {author} {\bibfnamefont {M.~B.}\
  \bibnamefont {Barbaro}}, \bibinfo {author} {\bibfnamefont {J.~A.}\
  \bibnamefont {Caballero}}, \ and\ \bibinfo {author} {\bibfnamefont {T.~W.}\
  \bibnamefont {Donnelly}},\ }\href {\doibase 10.1103/PhysRevC.90.035501}
  {\bibfield  {journal} {\bibinfo  {journal} {Phys. Rev.}\ }\textbf {\bibinfo
  {volume} {C90}},\ \bibinfo {pages} {035501} (\bibinfo {year} {2014})},\
  \Eprint {http://arxiv.org/abs/1407.8346} {arXiv:1407.8346 [nucl-th]}
  \BibitemShut {NoStop}%
\bibitem [{\citenamefont {Megias}\ \emph
  {et~al.}(2016{\natexlab{a}})\citenamefont {Megias}, \citenamefont {Amaro},
  \citenamefont {Barbaro}, \citenamefont {Caballero},\ and\ \citenamefont
  {Donnelly}}]{Megias:2016lke}%
  \BibitemOpen
  \bibfield  {author} {\bibinfo {author} {\bibfnamefont {G.~D.}\ \bibnamefont
  {Megias}}, \bibinfo {author} {\bibfnamefont {J.~E.}\ \bibnamefont {Amaro}},
  \bibinfo {author} {\bibfnamefont {M.~B.}\ \bibnamefont {Barbaro}}, \bibinfo
  {author} {\bibfnamefont {J.~A.}\ \bibnamefont {Caballero}}, \ and\ \bibinfo
  {author} {\bibfnamefont {T.~W.}\ \bibnamefont {Donnelly}},\ }\href {\doibase
  10.1103/PhysRevD.94.013012} {\bibfield  {journal} {\bibinfo  {journal} {Phys.
  Rev.}\ }\textbf {\bibinfo {volume} {D94}},\ \bibinfo {pages} {013012}
  (\bibinfo {year} {2016}{\natexlab{a}})},\ \Eprint
  {http://arxiv.org/abs/1603.08396} {arXiv:1603.08396 [nucl-th]} \BibitemShut
  {NoStop}%
\bibitem [{\citenamefont {Dolan}\ \emph {et~al.}(2020)\citenamefont {Dolan},
  \citenamefont {Megias},\ and\ \citenamefont {Bolognesi}}]{Dolan:2019bxf}%
  \BibitemOpen
  \bibfield  {author} {\bibinfo {author} {\bibfnamefont {S.}~\bibnamefont
  {Dolan}}, \bibinfo {author} {\bibfnamefont {G.~D.}\ \bibnamefont {Megias}}, \
  and\ \bibinfo {author} {\bibfnamefont {S.}~\bibnamefont {Bolognesi}},\ }\href
  {\doibase 10.1103/PhysRevD.101.033003} {\bibfield  {journal} {\bibinfo
  {journal} {Phys. Rev. D}\ }\textbf {\bibinfo {volume} {101}},\ \bibinfo
  {pages} {033003} (\bibinfo {year} {2020})},\ \Eprint
  {http://arxiv.org/abs/1905.08556} {arXiv:1905.08556 [hep-ex]} \BibitemShut
  {NoStop}%
\bibitem [{\citenamefont {Nieves}\ \emph {et~al.}(2011)\citenamefont {Nieves},
  \citenamefont {Ruiz~Simo},\ and\ \citenamefont
  {Vicente~Vacas}}]{Nieves:2011pp}%
  \BibitemOpen
  \bibfield  {author} {\bibinfo {author} {\bibfnamefont {J.}~\bibnamefont
  {Nieves}}, \bibinfo {author} {\bibfnamefont {I.}~\bibnamefont {Ruiz~Simo}}, \
  and\ \bibinfo {author} {\bibfnamefont {M.~J.}\ \bibnamefont
  {Vicente~Vacas}},\ }\href {\doibase 10.1103/PhysRevC.83.045501} {\bibfield
  {journal} {\bibinfo  {journal} {Phys. Rev.}\ }\textbf {\bibinfo {volume}
  {C83}},\ \bibinfo {pages} {045501} (\bibinfo {year} {2011})},\ \Eprint
  {http://arxiv.org/abs/1102.2777} {arXiv:1102.2777 [hep-ph]} \BibitemShut
  {NoStop}%
\bibitem [{\citenamefont {Abe}\ \emph {et~al.}(2020{\natexlab{a}})\citenamefont
  {Abe} \emph {et~al.}}]{T2K:2020jav}%
  \BibitemOpen
  \bibfield  {author} {\bibinfo {author} {\bibfnamefont {K.}~\bibnamefont
  {Abe}} \emph {et~al.} (\bibinfo {collaboration} {T2K}),\ }\href {\doibase
  10.1103/PhysRevD.101.112004} {\bibfield  {journal} {\bibinfo  {journal}
  {Phys. Rev. D}\ }\textbf {\bibinfo {volume} {101}},\ \bibinfo {pages}
  {112004} (\bibinfo {year} {2020}{\natexlab{a}})},\ \Eprint
  {http://arxiv.org/abs/2004.05434} {arXiv:2004.05434 [hep-ex]} \BibitemShut
  {NoStop}%
\bibitem [{\citenamefont {Abe}\ \emph {et~al.}(2020{\natexlab{b}})\citenamefont
  {Abe} \emph {et~al.}}]{T2K:2020sbd}%
  \BibitemOpen
  \bibfield  {author} {\bibinfo {author} {\bibfnamefont {K.}~\bibnamefont
  {Abe}} \emph {et~al.} (\bibinfo {collaboration} {T2K}),\ }\href {\doibase
  10.1103/PhysRevD.101.112001} {\bibfield  {journal} {\bibinfo  {journal}
  {Phys. Rev. D}\ }\textbf {\bibinfo {volume} {101}},\ \bibinfo {pages}
  {112001} (\bibinfo {year} {2020}{\natexlab{b}})},\ \Eprint
  {http://arxiv.org/abs/2002.09323} {arXiv:2002.09323 [hep-ex]} \BibitemShut
  {NoStop}%
\bibitem [{\citenamefont {Ruiz~Simo}\ \emph {et~al.}(2017)\citenamefont
  {Ruiz~Simo}, \citenamefont {Amaro}, \citenamefont {Barbaro}, \citenamefont
  {De~Pace}, \citenamefont {Caballero},\ and\ \citenamefont
  {Donnelly}}]{RuizSimo:2016rtu}%
  \BibitemOpen
  \bibfield  {author} {\bibinfo {author} {\bibfnamefont {I.}~\bibnamefont
  {Ruiz~Simo}}, \bibinfo {author} {\bibfnamefont {J.~E.}\ \bibnamefont
  {Amaro}}, \bibinfo {author} {\bibfnamefont {M.~B.}\ \bibnamefont {Barbaro}},
  \bibinfo {author} {\bibfnamefont {A.}~\bibnamefont {De~Pace}}, \bibinfo
  {author} {\bibfnamefont {J.~A.}\ \bibnamefont {Caballero}}, \ and\ \bibinfo
  {author} {\bibfnamefont {T.~W.}\ \bibnamefont {Donnelly}},\ }\href {\doibase
  10.1088/1361-6471/aa6a06} {\bibfield  {journal} {\bibinfo  {journal} {J.
  Phys. G}\ }\textbf {\bibinfo {volume} {44}},\ \bibinfo {pages} {065105}
  (\bibinfo {year} {2017})},\ \Eprint {http://arxiv.org/abs/1604.08423}
  {arXiv:1604.08423 [nucl-th]} \BibitemShut {NoStop}%
\bibitem [{\citenamefont {Ruiz~Simo}\ \emph {et~al.}(2016)\citenamefont
  {Ruiz~Simo}, \citenamefont {Amaro}, \citenamefont {Barbaro}, \citenamefont
  {De~Pace}, \citenamefont {Caballero}, \citenamefont {Megias},\ and\
  \citenamefont {Donnelly}}]{RuizSimo:2016ikw}%
  \BibitemOpen
  \bibfield  {author} {\bibinfo {author} {\bibfnamefont {I.}~\bibnamefont
  {Ruiz~Simo}}, \bibinfo {author} {\bibfnamefont {J.~E.}\ \bibnamefont
  {Amaro}}, \bibinfo {author} {\bibfnamefont {M.~B.}\ \bibnamefont {Barbaro}},
  \bibinfo {author} {\bibfnamefont {A.}~\bibnamefont {De~Pace}}, \bibinfo
  {author} {\bibfnamefont {J.~A.}\ \bibnamefont {Caballero}}, \bibinfo {author}
  {\bibfnamefont {G.~D.}\ \bibnamefont {Megias}}, \ and\ \bibinfo {author}
  {\bibfnamefont {T.~W.}\ \bibnamefont {Donnelly}},\ }\href {\doibase
  10.1016/j.physletb.2016.09.021} {\bibfield  {journal} {\bibinfo  {journal}
  {Phys. Lett.}\ }\textbf {\bibinfo {volume} {B762}},\ \bibinfo {pages} {124}
  (\bibinfo {year} {2016})},\ \Eprint {http://arxiv.org/abs/1607.08451}
  {arXiv:1607.08451 [nucl-th]} \BibitemShut {NoStop}%
\bibitem [{\citenamefont {Berger}\ and\ \citenamefont
  {Sehgal}(2007)}]{BSSPP2007}%
  \BibitemOpen
  \bibfield  {author} {\bibinfo {author} {\bibfnamefont {C.}~\bibnamefont
  {Berger}}\ and\ \bibinfo {author} {\bibfnamefont {L.~M.}\ \bibnamefont
  {Sehgal}},\ }\href {\doibase 10.1103/PhysRevD.76.113004} {\bibfield
  {journal} {\bibinfo  {journal} {Phys. Rev. D}\ }\textbf {\bibinfo {volume}
  {76}},\ \bibinfo {pages} {113004} (\bibinfo {year} {2007})}\BibitemShut
  {NoStop}%
\bibitem [{\citenamefont {Dytman}\ \emph {et~al.}(2021)\citenamefont {Dytman},
  \citenamefont {Hayato}, \citenamefont {Raboanary}, \citenamefont {Sobczyk},
  \citenamefont {Tena~Vidal},\ and\ \citenamefont
  {Vololoniaina}}]{Dytman:2021ohr}%
  \BibitemOpen
  \bibfield  {author} {\bibinfo {author} {\bibfnamefont {S.}~\bibnamefont
  {Dytman}}, \bibinfo {author} {\bibfnamefont {Y.}~\bibnamefont {Hayato}},
  \bibinfo {author} {\bibfnamefont {R.}~\bibnamefont {Raboanary}}, \bibinfo
  {author} {\bibfnamefont {J.~T.}\ \bibnamefont {Sobczyk}}, \bibinfo {author}
  {\bibfnamefont {J.}~\bibnamefont {Tena~Vidal}}, \ and\ \bibinfo {author}
  {\bibfnamefont {N.}~\bibnamefont {Vololoniaina}},\ }\href {\doibase
  10.1103/PhysRevD.104.053006} {\bibfield  {journal} {\bibinfo  {journal}
  {Phys. Rev. D}\ }\textbf {\bibinfo {volume} {104}},\ \bibinfo {pages}
  {053006} (\bibinfo {year} {2021})},\ \Eprint
  {http://arxiv.org/abs/2103.07535} {arXiv:2103.07535 [hep-ph]} \BibitemShut
  {NoStop}%
\bibitem [{\citenamefont {Niewczas}\ \emph {et~al.}(2021)\citenamefont
  {Niewczas}, \citenamefont {Nikolakopoulos}, \citenamefont {Sobczyk},
  \citenamefont {Jachowicz},\ and\ \citenamefont
  {Gonz\'alez-Jim\'enez}}]{Niewczas:2020fev}%
  \BibitemOpen
  \bibfield  {author} {\bibinfo {author} {\bibfnamefont {K.}~\bibnamefont
  {Niewczas}}, \bibinfo {author} {\bibfnamefont {A.}~\bibnamefont
  {Nikolakopoulos}}, \bibinfo {author} {\bibfnamefont {J.~T.}\ \bibnamefont
  {Sobczyk}}, \bibinfo {author} {\bibfnamefont {N.}~\bibnamefont {Jachowicz}},
  \ and\ \bibinfo {author} {\bibfnamefont {R.}~\bibnamefont
  {Gonz\'alez-Jim\'enez}},\ }\href {\doibase 10.1103/PhysRevD.103.053003}
  {\bibfield  {journal} {\bibinfo  {journal} {Phys. Rev. D}\ }\textbf {\bibinfo
  {volume} {103}},\ \bibinfo {pages} {053003} (\bibinfo {year} {2021})},\
  \Eprint {http://arxiv.org/abs/2011.05269} {arXiv:2011.05269 [hep-ph]}
  \BibitemShut {NoStop}%
\bibitem [{\citenamefont {Schwehr}\ \emph {et~al.}(2017)\citenamefont
  {Schwehr}, \citenamefont {Cherdack},\ and\ \citenamefont
  {Gran}}]{Valencia2p2hInGenie:2016}%
  \BibitemOpen
  \bibfield  {author} {\bibinfo {author} {\bibfnamefont {J.}~\bibnamefont
  {Schwehr}}, \bibinfo {author} {\bibfnamefont {D.}~\bibnamefont {Cherdack}}, \
  and\ \bibinfo {author} {\bibfnamefont {R.}~\bibnamefont {Gran}},\ }\href@noop
  {} {\  (\bibinfo {year} {2017})},\ \Eprint {http://arxiv.org/abs/1601.02038}
  {arXiv:1601.02038 [hep-ph]} \BibitemShut {NoStop}%
\bibitem [{\citenamefont {Amaro}\ \emph {et~al.}(2005)\citenamefont {Amaro},
  \citenamefont {Barbaro}, \citenamefont {Caballero}, \citenamefont {Donnelly},
  \citenamefont {Molinari},\ and\ \citenamefont {Sick}}]{Amaro:2004bs}%
  \BibitemOpen
  \bibfield  {author} {\bibinfo {author} {\bibfnamefont {J.~E.}\ \bibnamefont
  {Amaro}}, \bibinfo {author} {\bibfnamefont {M.~B.}\ \bibnamefont {Barbaro}},
  \bibinfo {author} {\bibfnamefont {J.~A.}\ \bibnamefont {Caballero}}, \bibinfo
  {author} {\bibfnamefont {T.~W.}\ \bibnamefont {Donnelly}}, \bibinfo {author}
  {\bibfnamefont {A.}~\bibnamefont {Molinari}}, \ and\ \bibinfo {author}
  {\bibfnamefont {I.}~\bibnamefont {Sick}},\ }\href {\doibase
  10.1103/PhysRevC.71.015501} {\bibfield  {journal} {\bibinfo  {journal} {Phys.
  Rev.}\ }\textbf {\bibinfo {volume} {C71}},\ \bibinfo {pages} {015501}
  (\bibinfo {year} {2005})},\ \Eprint {http://arxiv.org/abs/nucl-th/0409078}
  {arXiv:nucl-th/0409078 [nucl-th]} \BibitemShut {NoStop}%
\bibitem [{\citenamefont {Megias}(2017)}]{Megias:2017PhD}%
  \BibitemOpen
  \bibfield  {author} {\bibinfo {author} {\bibfnamefont {G.~D.}\ \bibnamefont
  {Megias}},\ }\emph {\bibinfo {title} {Charged-current neutrino interactions
  with nucleons and nuclei at intermediate energies}},\ \href {\doibase
  https://idus.us.es/xmlui/handle/11441/74826} {Ph.D. thesis},\ \bibinfo
  {school} {University of Seville, Spain} (\bibinfo {year} {2017}),\ \bibinfo
  {note} {https://idus.us.es/xmlui/handle/11441/74826}\BibitemShut {NoStop}%
\bibitem [{\citenamefont {Nikolakopoulos}\ \emph {et~al.}(2021)\citenamefont
  {Nikolakopoulos}, \citenamefont {Pandey}, \citenamefont {Spitz},\ and\
  \citenamefont {Jachowicz}}]{Nikolakopoulos:2020alk}%
  \BibitemOpen
  \bibfield  {author} {\bibinfo {author} {\bibfnamefont {A.}~\bibnamefont
  {Nikolakopoulos}}, \bibinfo {author} {\bibfnamefont {V.}~\bibnamefont
  {Pandey}}, \bibinfo {author} {\bibfnamefont {J.}~\bibnamefont {Spitz}}, \
  and\ \bibinfo {author} {\bibfnamefont {N.}~\bibnamefont {Jachowicz}},\ }\href
  {\doibase 10.1103/PhysRevC.103.064603} {\bibfield  {journal} {\bibinfo
  {journal} {Phys. Rev. C}\ }\textbf {\bibinfo {volume} {103}},\ \bibinfo
  {pages} {064603} (\bibinfo {year} {2021})},\ \Eprint
  {http://arxiv.org/abs/2010.05794} {arXiv:2010.05794 [nucl-th]} \BibitemShut
  {NoStop}%
\bibitem [{\citenamefont {Megias}\ \emph
  {et~al.}(2016{\natexlab{b}})\citenamefont {Megias}, \citenamefont {Amaro},
  \citenamefont {Barbaro}, \citenamefont {Caballero},\ and\ \citenamefont
  {Donnelly}}]{Megias:2016ee}%
  \BibitemOpen
  \bibfield  {author} {\bibinfo {author} {\bibfnamefont {G.~D.}\ \bibnamefont
  {Megias}}, \bibinfo {author} {\bibfnamefont {J.~E.}\ \bibnamefont {Amaro}},
  \bibinfo {author} {\bibfnamefont {M.~B.}\ \bibnamefont {Barbaro}}, \bibinfo
  {author} {\bibfnamefont {J.~A.}\ \bibnamefont {Caballero}}, \ and\ \bibinfo
  {author} {\bibfnamefont {T.~W.}\ \bibnamefont {Donnelly}},\ }\href@noop {}
  {\bibfield  {journal} {\bibinfo  {journal} {Phys. Rev. D}\ }\textbf {\bibinfo
  {volume} {94}},\ \bibinfo {pages} {013012} (\bibinfo {year}
  {2016}{\natexlab{b}})}\BibitemShut {NoStop}%
\bibitem [{\citenamefont {Gonz\'alez-Jim\'enez}\ \emph
  {et~al.}(2020)\citenamefont {Gonz\'alez-Jim\'enez}, \citenamefont {Barbaro},
  \citenamefont {Caballero}, \citenamefont {Donnelly}, \citenamefont
  {Jachowicz}, \citenamefont {Megias}, \citenamefont {Niewczas}, \citenamefont
  {Nikolakopoulos},\ and\ \citenamefont
  {Ud\'\i{}as}}]{Gonzalez-Jimenez:2019ejf}%
  \BibitemOpen
  \bibfield  {author} {\bibinfo {author} {\bibfnamefont {R.}~\bibnamefont
  {Gonz\'alez-Jim\'enez}}, \bibinfo {author} {\bibfnamefont {M.~B.}\
  \bibnamefont {Barbaro}}, \bibinfo {author} {\bibfnamefont {J.~A.}\
  \bibnamefont {Caballero}}, \bibinfo {author} {\bibfnamefont {T.~W.}\
  \bibnamefont {Donnelly}}, \bibinfo {author} {\bibfnamefont {N.}~\bibnamefont
  {Jachowicz}}, \bibinfo {author} {\bibfnamefont {G.~D.}\ \bibnamefont
  {Megias}}, \bibinfo {author} {\bibfnamefont {K.}~\bibnamefont {Niewczas}},
  \bibinfo {author} {\bibfnamefont {A.}~\bibnamefont {Nikolakopoulos}}, \ and\
  \bibinfo {author} {\bibfnamefont {J.~M.}\ \bibnamefont {Ud\'\i{}as}},\ }\href
  {\doibase 10.1103/PhysRevC.101.015503} {\bibfield  {journal} {\bibinfo
  {journal} {Phys. Rev. C}\ }\textbf {\bibinfo {volume} {101}},\ \bibinfo
  {pages} {015503} (\bibinfo {year} {2020})},\ \Eprint
  {http://arxiv.org/abs/1909.07497} {arXiv:1909.07497 [nucl-th]} \BibitemShut
  {NoStop}%
\bibitem [{\citenamefont {Megias}\ \emph {et~al.}(2015)\citenamefont {Megias}
  \emph {et~al.}}]{Megias:2014qva}%
  \BibitemOpen
  \bibfield  {author} {\bibinfo {author} {\bibfnamefont {G.~D.}\ \bibnamefont
  {Megias}} \emph {et~al.},\ }\href {\doibase 10.1103/PhysRevD.91.073004}
  {\bibfield  {journal} {\bibinfo  {journal} {Phys. Rev.}\ }\textbf {\bibinfo
  {volume} {D91}},\ \bibinfo {pages} {073004} (\bibinfo {year} {2015})},\
  \Eprint {http://arxiv.org/abs/1412.1822} {arXiv:1412.1822 [nucl-th]}
  \BibitemShut {NoStop}%
\bibitem [{\citenamefont {Abe}\ \emph {et~al.}(2013)\citenamefont {Abe} \emph
  {et~al.}}]{Abe:2012av}%
  \BibitemOpen
  \bibfield  {author} {\bibinfo {author} {\bibfnamefont {K.}~\bibnamefont
  {Abe}} \emph {et~al.} (\bibinfo {collaboration} {T2K}),\ }\href {\doibase
  10.1103/PhysRevD.87.012001, 10.1103/PhysRevD.87.019902} {\bibfield  {journal}
  {\bibinfo  {journal} {Phys. Rev.}\ }\textbf {\bibinfo {volume} {D87}},\
  \bibinfo {pages} {012001} (\bibinfo {year} {2013})},\ \bibinfo {note}
  {[Addendum: Phys. Rev.D87,no.1,019902(2013)]},\ \Eprint
  {http://arxiv.org/abs/1211.0469} {arXiv:1211.0469 [hep-ex]} \BibitemShut
  {NoStop}%
\bibitem [{t2k(2016)}]{t2kfluxurl}%
  \BibitemOpen
  \href {\doibase 10.5281/zenodo.5734267} {\enquote {\bibinfo {title} {Neutrino
  beam flux prediction 2016},}\ }\bibinfo {howpublished} {\url{
  https://doi.org/10.5281/zenodo.5734267}} (\bibinfo {year} {2016}),\ \bibinfo
  {note} {accessed: 2019-08-07}\BibitemShut {NoStop}%
\bibitem [{\citenamefont {Martini}\ \emph {et~al.}(2016)\citenamefont
  {Martini}, \citenamefont {Jachowicz}, \citenamefont {Ericson}, \citenamefont
  {Pandey}, \citenamefont {Van~Cuyck},\ and\ \citenamefont
  {Van~Dessel}}]{Martini:2016eec}%
  \BibitemOpen
  \bibfield  {author} {\bibinfo {author} {\bibfnamefont {M.}~\bibnamefont
  {Martini}}, \bibinfo {author} {\bibfnamefont {N.}~\bibnamefont {Jachowicz}},
  \bibinfo {author} {\bibfnamefont {M.}~\bibnamefont {Ericson}}, \bibinfo
  {author} {\bibfnamefont {V.}~\bibnamefont {Pandey}}, \bibinfo {author}
  {\bibfnamefont {T.}~\bibnamefont {Van~Cuyck}}, \ and\ \bibinfo {author}
  {\bibfnamefont {N.}~\bibnamefont {Van~Dessel}},\ }\href {\doibase
  10.1103/PhysRevC.94.015501} {\bibfield  {journal} {\bibinfo  {journal} {Phys.
  Rev. C}\ }\textbf {\bibinfo {volume} {94}},\ \bibinfo {pages} {015501}
  (\bibinfo {year} {2016})},\ \Eprint {http://arxiv.org/abs/1602.00230}
  {arXiv:1602.00230 [nucl-th]} \BibitemShut {NoStop}%
\bibitem [{\citenamefont {Pandey}\ \emph {et~al.}(2014)\citenamefont {Pandey},
  \citenamefont {Jachowicz}, \citenamefont {Ryckebusch}, \citenamefont
  {Van~Cuyck},\ and\ \citenamefont {Cosyn}}]{Pandey:2013cca}%
  \BibitemOpen
  \bibfield  {author} {\bibinfo {author} {\bibfnamefont {V.}~\bibnamefont
  {Pandey}}, \bibinfo {author} {\bibfnamefont {N.}~\bibnamefont {Jachowicz}},
  \bibinfo {author} {\bibfnamefont {J.}~\bibnamefont {Ryckebusch}}, \bibinfo
  {author} {\bibfnamefont {T.}~\bibnamefont {Van~Cuyck}}, \ and\ \bibinfo
  {author} {\bibfnamefont {W.}~\bibnamefont {Cosyn}},\ }\href {\doibase
  10.1103/PhysRevC.89.024601} {\bibfield  {journal} {\bibinfo  {journal} {Phys.
  Rev. C}\ }\textbf {\bibinfo {volume} {89}},\ \bibinfo {pages} {024601}
  (\bibinfo {year} {2014})},\ \Eprint {http://arxiv.org/abs/1310.6885}
  {arXiv:1310.6885 [nucl-th]} \BibitemShut {NoStop}%
\bibitem [{\citenamefont {Stowell}\ \emph {et~al.}(2017)\citenamefont {Stowell}
  \emph {et~al.}}]{Stowell:2016jfr}%
  \BibitemOpen
  \bibfield  {author} {\bibinfo {author} {\bibfnamefont {P.}~\bibnamefont
  {Stowell}} \emph {et~al.},\ }\href {\doibase 10.1088/1748-0221/12/01/P01016}
  {\bibfield  {journal} {\bibinfo  {journal} {JINST}\ }\textbf {\bibinfo
  {volume} {12}},\ \bibinfo {pages} {P01016} (\bibinfo {year} {2017})},\
  \Eprint {http://arxiv.org/abs/1612.07393} {arXiv:1612.07393 [hep-ex]}
  \BibitemShut {NoStop}%
\bibitem [{\citenamefont {Ruterbories}\ \emph {et~al.}(2022)\citenamefont
  {Ruterbories} \emph {et~al.}}]{MINERvA:2022mnw}%
  \BibitemOpen
  \bibfield  {author} {\bibinfo {author} {\bibfnamefont {D.}~\bibnamefont
  {Ruterbories}} \emph {et~al.} (\bibinfo {collaboration} {MINERvA}),\ }\href
  {\doibase 10.1103/PhysRevLett.129.021803} {\bibfield  {journal} {\bibinfo
  {journal} {Phys. Rev. Lett.}\ }\textbf {\bibinfo {volume} {129}},\ \bibinfo
  {pages} {021803} (\bibinfo {year} {2022})},\ \Eprint
  {http://arxiv.org/abs/2203.08022} {arXiv:2203.08022 [hep-ex]} \BibitemShut
  {NoStop}%
\bibitem [{\citenamefont {Van~Dessel}\ \emph {et~al.}(2018)\citenamefont
  {Van~Dessel}, \citenamefont {Jachowicz}, \citenamefont
  {Gonz\'alez-Jim\'enez}, \citenamefont {Pandey},\ and\ \citenamefont
  {Van~Cuyck}}]{VanDessel:2017ery}%
  \BibitemOpen
  \bibfield  {author} {\bibinfo {author} {\bibfnamefont {N.}~\bibnamefont
  {Van~Dessel}}, \bibinfo {author} {\bibfnamefont {N.}~\bibnamefont
  {Jachowicz}}, \bibinfo {author} {\bibfnamefont {R.}~\bibnamefont
  {Gonz\'alez-Jim\'enez}}, \bibinfo {author} {\bibfnamefont {V.}~\bibnamefont
  {Pandey}}, \ and\ \bibinfo {author} {\bibfnamefont {T.}~\bibnamefont
  {Van~Cuyck}},\ }\href {\doibase 10.1103/PhysRevC.97.044616} {\bibfield
  {journal} {\bibinfo  {journal} {Phys. Rev. C}\ }\textbf {\bibinfo {volume}
  {97}},\ \bibinfo {pages} {044616} (\bibinfo {year} {2018})},\ \Eprint
  {http://arxiv.org/abs/1704.07817} {arXiv:1704.07817 [nucl-th]} \BibitemShut
  {NoStop}%
\end{thebibliography}%

\end{document}